\documentclass[english,15pt]{article}
\usepackage[T1]{fontenc}
\usepackage[latin1]{inputenc}
\usepackage{geometry}
\usepackage{float}
\usepackage{mathrsfs}
\usepackage{amsmath}
\usepackage{amssymb}
\usepackage{graphicx}
\usepackage{hyperref}
\usepackage[color=yellow]{todonotes}
\hypersetup{
    colorlinks=true,
    linkcolor=blue,
    filecolor=magenta,      
    urlcolor=cyan,
    citecolor = red,
}

\usepackage{yhmath}
\makeatletter

\usepackage{bm}
\numberwithin{equation}{section}

\floatstyle{ruled}
\usepackage{algorithm}
\usepackage{algpseudocode}
\usepackage{caption}
\usepackage{needspace}

\DeclareCaptionFormat{algor}{%
  \needspace{6\baselineskip}\hrulefill\par\offinterlineskip\vskip1pt%
    \textbf{#1#2:} #3\offinterlineskip\hrulefill}
\DeclareCaptionStyle{algori}{singlelinecheck=off,format=algor,labelsep=space}

\usepackage{algorithm}
 \usepackage{algorithmicx}

\usepackage{amsthm}

\usepackage{mathrsfs}
\usepackage{amssymb}

\usepackage{amsfonts}

\usepackage{epsfig}

\usepackage{bm}

\usepackage{mathrsfs}

\usepackage{enumerate}

\@ifundefined{definecolor}{\@ifundefined{definecolor}
 {\@ifundefined{definecolor}
 {\usepackage{color}}{}
}{}
}{}
\newtheorem{ass}{Assumption}[section]

\usepackage{subfig}\usepackage[all]{xy}

\newtheorem{prop}{Proposition}[section]

\newcounter{hypA}

\newcounter{hypB}

\newcounter{hypD}

\usepackage[nameinlink, capitalise]{cleveref}

\providecommand{\algorithmname}{Algorithm}
\floatname{algorithm}{\protect\algorithmname}

\usepackage{babel}\date{}

\makeatother

\usepackage{diagbox}

\begin{document}

\begin{center}

{\Large \textbf{Local Sequential MCMC for Data Assimilation with Applications in Geoscience}}

\vspace{0.5cm}

Hamza Ruzayqat and Omar Knio\\
\smallskip
{\footnotesize Applied Mathematics and Computational Science Program, \\ Computer, Electrical and Mathematical Sciences and Engineering Division, \\ King Abdullah University of Science and Technology, Thuwal, 23955-6900, KSA.} \\
{\footnotesize E-Mail:\,}  \texttt{\emph{\footnotesize hamza.ruzayqat@kaust.edu.sa}},  \texttt{\emph{\footnotesize omar.knio@kaust.edu.sa}} 

\begin{abstract}
This paper presents a new data assimilation (DA) scheme based on a sequential Markov Chain Monte Carlo (SMCMC) DA technique \cite{smcmc} which is provably convergent and has been recently used for filtering, particularly for high-dimensional non-linear, and potentially, non-Gaussian state-space models. Unlike particle filters, which can be considered exact methods and can be used for filtering non-linear, non-Gaussian models, SMCMC does not assign weights to the samples/particles, and therefore, the method does not suffer from the issue of weight-degeneracy when a relatively small number of samples is used. We design a localization approach within the SMCMC framework that focuses on regions where observations are located and restricts the transition densities included in the filtering distribution of the state to these regions. This results in immensely reducing the effective degrees of freedom and thus improving the efficiency. We test the new technique on high-dimensional ($d \sim 10^4 - 10^5$) linear Gaussian model and non-linear shallow water models with Gaussian noise with real and synthetic observations. For two of the numerical examples, the observations mimic the data generated by the Surface Water and Ocean Topography (SWOT) mission led by NASA, which is a swath of ocean height observations that changes location at every assimilation time step. We also use a set of ocean drifters' real observations in which the drifters are moving according the ocean kinematics and assumed to have uncertain locations at the time of assimilation. We show that when higher accuracy is required, the proposed algorithm is superior in terms of efficiency and accuracy over competing ensemble methods and the original SMCMC filter.

\bigskip

\noindent \textbf{Keywords}: Data Assimilation, Localization, Sequential Markov Chain Monte Carlo, High-Dimensional Filtering.
\\
\noindent \textbf{MSC classes}:	62M20, 60G35, 60J20, 94A12, 93E11, 65C40
\\
\noindent \textbf{Corresponding author}: Hamza Ruzayqat. E-mail:
\href{mailto:hamza.ruzayqat@kaust.edu.sa}{hamza.ruzayqat@kaust.edu.sa}  \\
\noindent \textbf{Code:} \url{https://github.com/ruzayqat/LSMCMC}
\end{abstract}
\end{center}

\section{Introduction}
Stochastic filtering or data assimilation (DA) plays a crucial role in understanding and predicting complex systems states as accurately as possible. By combining partial, noisy observational data with numerical models, DA aims to predict the conditional probability distribution, or the filter distribution, of underlying hidden state variables that cannot be directly observed. In practice, data assimilation is widely utilized across various fields. In finance, it helps in predicting market trends and managing risks by improving the accuracy of financial models. In engineering, it is used to optimize systems and processes by refining predictions based on real-time data. In geophysical sciences, data assimilation is crucial for applications such as numerical weather prediction, where it provides vital information for forecasting weather patterns, understanding atmospheric conditions, and modeling oceanographic phenomena. These applications are particularly significant for climate science, where data assimilation aids in tracking climate change, ensuring maritime safety, and predicting natural hazards like tsunamis \cite{tsunami}. For a comprehensive introduction on DA and stochastic filtering see \cite{bain,cappe}. For a deeper exploration of specific applications and methodologies see \cite{carrassi, ghil, kalnay}.

Ensemble methods, such as the ensemble Kalman filter (EnKF) \cite{evensen1, houtekamer, evensen2} and sequential Monte Carlo methods such as particle filters (PFs) \cite{PF1, delm13, leeuwen}, are powerful DA tools that have gained significant attention in recent work; see \cite{vetra} for a review of DA methods. These probabilistic methods, which can be formalized within a sequential Bayesian inference framework, use multiple ensemble members or particles to represent and propagate the system's state dynamics, in conjunction with the available observations. Despite their inherent computational efficiency and their wide usage, ensemble Kalman methods can be inaccurate when dealing with strongly nonlinear and non-Gaussian models. Additionally, they have a tendency to significantly underestimate the overall uncertainty in the system \cite{pires, storto, vetra}. PFs, on the other hand, are exact methods that assign weights to the ensemble members (also called particles) that reflect the likelihood of the observations. As the number of particles approaches infinity, the PF estimates the conditional distribution of the hidden state exactly, regardless of the type of noise present in the system or whether the underlying dynamical model is linear or non-linear. The computational complexity of PFs is fixed per-time-step update. However, in order to prevent the so-called weight degeneracy (e.g. \cite{bengtsson}), the number of particles must grow exponentially with the dimensionality of the state space. This is the primary limitation of standard PF approaches, as they are typically only practical for low-dimensional problems. Several works in the literature have explored different techniques such as tempering, lagging and nudging to tackle this issue; see e.g. \cite{ades, laggedPF, cotter, kantas}. 

Another exact method for filtering is Markov chain Monte Carlo (MCMC) \cite{cappe}. MCMC can be used to sample from the filter distribution, however, its computational complexity grows linearly with time, which makes it impractical for real-time applications. A possible solution to the linear growth of complexity is to use sequential MCMC. At each assimilation time step, one builds an MCMC chain initialized from one random sample at the previous time step. Sequential MCMC has been used in dynamic contexts in several works. For instance, in the work of \cite{septier}, a sequential MCMC based on Hamiltonian Monte Carlo (HMC) was presented. The authors use HMC to build a chain of $N$ samples from the posterior initialized from a uniformly picked sample from the previous time step. Another example is the work of \cite{carmi}, where the authors use Sequential MCMC for dynamical cluster tracking.
 
In \cite{smcmc}, a more practical filtering approach is presented where sequential MCMC chains are used to target the filter distribution at each time-step update. The method was mainly developed for state-space models (SSMs) with unknown random observational locations, but it can be used for general SSMs. The initial idea behind their scheme first appeared in \cite{berzuini} and was later used for point process filtering in \cite{martin}.  The authors in \cite{smcmc} demonstrate that when high accuracy is needed and the state dimension is very large, sequential MCMC filtering scheme (which we refer to as SMCMC from now on) is much more accurate and efficient when compared to ensemble methods. The computational complexity of the algorithm ranges from $O(Nd)$ to $O(Nd^2)$ depending on the structure of the hidden state noise covariance matrix, where $N$ is the number of samples used in the MCMC chains and $d$ is the hidden state dimension.

In many real-world applications, observations on the system are either very sparse in the domain of interest or localized in particular regions. Many localization techniques have been developed in the literature to leverage this, thereby improving the overall performance. In \cite{gaspari}, the authors introduced a distance-based tapering function which is compactly supported that can be multiplied by the covariance matrix or its estimation to reduce the influence of observations that are spatially far from the state variables being updated. This type of techniques have been largely used in the literature for improving ensemble methods performance, such as EnKF or ensemble transform Kalman filter; see e.g. \cite{houtekamer, hamill, ott, miyoshi}. Whereas fixed localization functions have been widely used, it was shown that adaptive localization, which dynamically adjusts the localization radius length scale based on the spatial correlation structure, is very effective in practice, e.g. \cite{anderson}. Furthermore, localization and adaptive localization have been used in PFs as well. An approach is to apply localization directly to the weights and resampling process aiming to reduce the path degeneracy and improving the effective sample size. Examples are the works of \cite{poterjoy, farchi}, where localization is applied to the particles update ensuring that only observations within a distance influence each particle.

In this work, we build upon the SMCMC approach in \cite{smcmc} and propose a localized variant. The idea is based upon the hypothesis that, when the domain is divided up into local subdomains of moderate size, then one can use an approximate state error covariance matrix by restricting the original covariance matrix to a subspace whose dimension may be substantially lower than $d$. This subspace is mainly focused on regions where observations exist. Since the main cost of SMCMC filter comes from evaluating pointwisely the transition and likelihood densities in an approximation to the filtering density, which usually involves vector-matrix multiplications, the proposed method aims at dramatically reducing this cost while still maintaining a higher accuracy. The cost order of the new algorithm ranges from $O(Nd')$ to $O(Nd'^2)$ with $d' \ll d$. 

The contributions of this article are as follows:
\begin{itemize}
\item We introduce a localization technique to the SMCMC filter based on using an approximation of the covariance matrix of the hidden state's error.
\item We apply the method on several challenging SSMs including rotating shallow water equations (RSWEs) and show that, for a fixed accuracy, the cost of the proposed method is 5 to 10 times less versus SMCMC, EnKF and localized EnKF (LEnKF).

\item We demonstrate the methodology on filtering problems with two types of observations. 
\begin{enumerate}[(a)]
\item The first is synthetic swaths of data that move in time in a cyclic manner mirroring the data produced by SWOT mission led by NASA in collaboration with international partners \cite{nasa, swot}, which is not available for public yet. 

\item The second set is obtained from the National Oceanic and Atmospheric Administration (NOAA) and contains observations from ocean drifters in a domain of the Atlantic Ocean \cite{noaa, adam1}.
\end{enumerate}
\end{itemize}

This article is structured as follows.  \autoref{sec:model_prel} describes the class of SSMs considered in this paper. In \autoref{sec:smcmc} we provide a brief review of the SMCMC filter. \autoref{sec:lsmcmc} proposes the localization SMCMC algorithm. In \autoref{sec:numerics} we demonstrate the methodology on several SSMs. Finally, in \autoref{sec:conclusion} we conclude with a summary of key results and a discussion of possible future directions. 

\section{Preliminaries}\label{sec:model}
\label{sec:model_prel}
Consider a two-dimensional grid $\mathsf{G}$ of size $N_g\in \mathbb{N}$. $\mathsf{G}$ is typically a connected region such as a square or a rectangle with $N_g = N_g^x \times N_g^y$, where $N_g^x, N_g^y \in \mathbb{N}$. Assume there is an unknown continuous time stochastic process $(\mathbf{Z}_t)_{t\geq 0}$, with $\mathbf{Z}_t\in\mathbb{R}^d$, and $d\in \mathbb{N}$ is a multiple of $N_g^x N_g^y$. We will consider noisy observations that are arriving at times, $\mathsf{T}:=\{t_1,t_2,\dots\}\subset \mathbb{N}$, $0<t_1<t_2<\cdots$, at a collection of known locations. Let $m_k\in \mathbb{N}$ be the number of observed locations at time $t_k$. Then, at the observed location $j\in \{1, \cdots, m_k\}$, the observation is denoted by $Y_{t_k}^{(j)} \in \mathbb{R}^{s}$, $s\in \mathbb{N}$ the number of observed variables per observational location. We denote by $\mathbf{Y}_{t_k}\in \mathbb{R}^{d_y^k}$, $d_y^k = sm_k$, the collection of observations at time $t_k$. We have a superscript $k$ attached to $d_y$ to indicate that the number of available observations during the assimilation process may not be fixed. We will assume the time-lag between two consecutive observations to be $L\in \mathbb{N}$, that is $t_{k+1} - t_{k}=L$.

The modeling approach for the process $\mathbf{Z}_t$ is based on a continuous-time framework, motivated by applications in atmospheric and ocean sciences. In these domains, physical quantities such as height, wind, or water velocity are described using continuous-time, spatially-varying physical models represented by systems of partial differential equations (PDEs). To account for model uncertainty, the dynamics of $\mathbf{Z}_t$ are modeled using a stochastic process, where the noise can be incorporated either continuously (as in stochastic PDEs) or discretely in time.

The framework requires that at the discrete time instances in the set $\mathsf{T}$, the discrete-time process $(\mathbf{Z}_{t_k})_{k\geq 1}$ forms a Markov process with a known and well-defined (positive) transition density. Specifically, for any set $A \subseteq \mathbb{R}^d$, the transition dynamics of $(\mathbf{Z}_t)_{t\geq 0}$ are given by:
\begin{equation}
\label{eq:signal_model}
\mathbb{P}(\mathbf{Z}_{t_k}\in A|\mathbf{z}_{t_{k-1}}) = \int_A f_k(\mathbf{z}_{t_{k-1}},\mathbf{z}_{t_k}) d\mathbf{z}_{t_k}
\end{equation}
where $k \in \mathbb{N}$, $\mathbb{P}$ denotes probability, and the subscript $k$ in the transition density $f_k$ accounts for possible time-inhomogeneity of the process $(\mathbf{Z}_t)_{t\geq 0}$ or dependence on the time increments. The initial state $\mathbf{Z}_0$ is assumed to be known.

We will assume that each observation vector $\mathbf{Y}_{t_k}$ depends only on $\mathbf{Z}_{t_k}$ and there is a positive conditional likelihood density, i.e. for $k\in\mathbb{N}$, 
$B\subseteq \mathbb{R}^{d_y}$,
\begin{equation}
\label{eq:obs_model}
\mathbb{P}\left(\mathbf{Y}_{t_k}\in B|(\mathbf{Z}_t)_{t\geq 0},(\mathbf{Y}_{t})_{t\in\mathsf{T}\setminus\{t_k\}}\right) =\int_B g_k(\mathbf{z}_{t_k},\mathbf{y}_{t_k}) d\mathbf{y}_{t_k}
\end{equation}
where $g_k$ denotes the observations likelihood density at time $t_k$. The subscript is included in $g_k$ to account for possible time-inhomogeneous structure of $(\mathbf{Y}_t)_{t\in \mathsf{T}}$.

We will assume that the transition density $f_k$ (or a suitable approximation of it) can be evaluated pointwisely. This includes examples such as stochastic differential equations (SDEs) and their various time discretization approximations, as well as stochastic PDEs or PDEs with discrete-time additive noise. These types of models provide a framework for specifying the state-space dynamics of the process of interest. A SSM that fits within this general setup and applies to all examples considered in the article is presented below.

\textbf{Hidden signal:}
We will consider $\mathbf{Z}_t$ to be a vector containing hidden state variables at positions defined on $\mathsf{G}$. Let $\Phi:\mathbb{R}^d \times \mathbb{R}_+^2 \to \mathbb{R}^d$ be some linear or nonlinear function defined on its domain, then for $k\in \mathbb{N}$:
\begin{equation}
\label{eq:signal_unobserved}
\mathbf{Z}_t  = \Phi(\mathbf{Z}_{t_{k-1}},t_{k-1}; t), \qquad t \in (t_{k-1},t_k),
\end{equation}
and at time of observation $t_k$:
\begin{equation}
\label{eq:signal_observed}
\mathbf{Z}_{t_k} = \Phi(\mathbf{Z}_{t_{k-1}}, t_{k-1}; t_k) + \mathbf{W}_{t_k},
\end{equation}
where $\mathbf{W}_{t_k}\stackrel{\textrm{\emph{i.i.d.}}}{\sim}\mathcal{N}_d(0,Q_k)$ is an i.i.d.~sequence of $d-$dimensional Gaussian random variables of zero mean and covariance matrix $Q_k$. 

\textbf{Observations:} In addition, the following observational model is considered
\begin{equation}
\label{eq:obs_model}
\mathbf{Y}_{t_k} = \mathscr{O}_{t_k}(\mathbf{Z}_{t_k}) + \mathbf{V}_{t_k},
\end{equation}
where $\mathscr{O}_{t_k}:\mathbb{R}^d\to \mathbb{R}^{d_y^k}$ is an $\mathbb{R}^{d_y^k}$-vector valued function and $\mathbf{V}_{t_k} \stackrel{\textrm{i.i.d.}}{\sim} \mathcal{N}_{d_y^k}(0,R_k)$ is an i.i.d.~sequence of $d_y^k-$dimensional Gaussian random variables of zero mean and covariance matrix $R_k$.

\section{SMCMC Filter}
\label{sec:smcmc}
Inference of the hidden state is performed using conditional distributions given the available observations. One can either consider the whole path trajectory 
\begin{equation*}
\mathbb{P}(\mathbf{Z}_{t_1},\ldots,\mathbf{Z}_{t_k}|(\mathbf{Y}_{t_p})_{p\leq {k}}) \qquad \text{(smoothing)}
\end{equation*}
or just the marginal
\begin{equation*}
\mathbb{P}(\mathbf{Z}_{t_k}|(\mathbf{Y}_{t_p})_{p\leq {k}}) \qquad \text{(filtering)}.
\end{equation*}
For $k\in\mathbb{N}$ we define the smoothing density (the dependence on observations is dropped from the notation)
\begin{align}
\label{eq:smoothing_density}
\Pi_k(\mathbf{z}_{t_{1}},\ldots, \mathbf{z}_{t_k}) &\propto \prod_{p=1}^k f_p(\mathbf{z}_{t_{p-1}},\mathbf{z}_{t_p}) g_p(\mathbf{z}_{t_p},\mathbf{y}_{t_p}) \nonumber\\
&\propto f_k(\mathbf{z}_{t_{k-1}},\mathbf{z}_{t_k}) g_k(\mathbf{z}_{t_k},\mathbf{y}_{t_k}) \Pi_{k-1}(\mathbf{z}_{t_{1:k-1}}). 
\end{align}
Let $\pi_k(\mathbf{z}_{t_k})$ be the marginal in the $\mathbf{z}_{t_k}$ coordinate of the smoothing distribution $\Pi_k$. It easily follows that the filter density can be obtained recursively
\begin{equation}
\label{eq:filter_recursion}
\pi_k(\mathbf{z}_{t_k}) \propto g_k(\mathbf{z}_{t_k},\mathbf{y}_{t_k})\int_{\mathbb{R}^d}f_k(\mathbf{z}_{t_{k-1}},\mathbf{z}_{t_k})\pi_{k-1}(\mathbf{z}_{t_{k-1}})d\mathbf{z}_{t_{k-1}}.    
\end{equation}
The filtering problem we are considering is inherently discrete-time in nature, as the observations are obtained at discrete time instances. However, we can still obtain the conditional probability distribution $\mathbb{P}(\mathbf{Z}_t | (\mathbf{Y}_{t_p})_{p \leq k})$ for any time $t\in (t_k, t_{k+1})$. This can be achieved by integrating the previous filter distribution $\pi_{k-1}$ with the corresponding transition density, and then applying the Chapman-Kolmogorov equations. This allows us to propagate the filter distribution continuously in time, even though the observations are discrete.

In \cite{smcmc}, the authors propose a method to efficiently approximate \eqref{eq:filter_recursion} directly via a sequence of MCMC chains initialized from a previously obtained approximation of $\pi_{k-1}$. 
At time $t_1$ their algorithm targets $\pi_1(\mathbf{z}_{t_1})$ exactly by running an MCMC kernel with invariant distribution $\pi_1(\mathbf{z}_{t_1})$ for $N$ steps, that is to draw $N$ samples from
\begin{equation}
\label{eq:pi_1}
\pi_1(\mathbf{z}_{t_1}) \propto g_1(\mathbf{z}_{t_1},\mathbf{y}_{t_1})f_1(\mathbf{Z}_0,\mathbf{z}_{t_1}).
\end{equation}
At later times $t_k$, $k\geq 2$, it targets an approximation of the filter distribution obtained by replacing $\pi_{k-1}(\mathbf{z}_{t_{k-1}})$ in \eqref{eq:filter_recursion} by an empirical density 
\begin{equation*}
S_{k-1}^N(\mathbf{z}_{t_{k-1}}):=\frac{1}{N} \sum_{i=1}^N \textcolor{black}{\delta_{\left\{\mathbf{Z}_{t_{k-1}}^{(i)}\right\}}}(\mathbf{z}_{t_{k-1}}),
\end{equation*}
where \textcolor{black}{$\delta_{\left\{\mathbf{Z}_{t_k}\right\}}(\mathbf{z})$ is the Dirac delta measure centered at $\mathbf{Z}_{t_k}$}, and $\mathbf{Z}_{t_{k-1}}^{(1)}, \mathbf{Z}_{t_{k-1}}^{(2)}, \cdots, \mathbf{Z}_{t_{k-1}}^{(N)}$ are the MCMC samples obtained in the preceding time step with the superscripts denoting MCMC iteration number. Plugging $S_{k-1}^N(\mathbf{z}_{t_{k-1}})$ in \eqref{eq:filter_recursion}, we obtain the approximation
\begin{equation*}
\pi_k^N (\mathbf{z}_{t_k})\propto g_k(\mathbf{z}_{t_k},\mathbf{y}_{t_k}) \frac{1}{N}\sum_{i=1}^N f_k(\mathbf{Z}_{t_{k-1}}^{(i)}, \mathbf{z}_{t_k} ).
\end{equation*}
Note that this distribution is just a marginal of the joint distribution with density
\begin{align}
\label{eq:filter_approx}
\pi_k^{N}(\mathbf{z}_{t_k},j) \propto g_k(\mathbf{z}_{t_k},\mathbf{y}_{t_k}) f_k(\mathbf{Z}_{t_{k-1}}^{(j)},\mathbf{z}_{t_k})p(j)
\end{align}
where $p(j)$ is the uniform distribution over the integers $\{1,\cdots, N\}$ so that $\pi_k^{N}(\mathbf{z}_{t_k},j)$ admits $\pi_k^N(\mathbf{z}_{t_k})$ as its marginal. The idea is to sample from $\pi_k^{N}(\mathbf{z}_{t_k},j)$ using an MCMC kernel. In this paper, we use a random walk Metropolis (RWM) kernel. RWM is a commonly used and versatile choice of MCMC kernels. In the numerical implementations presented later, RWM proved to be effective and efficient. 

For a given function $\varphi:\mathbb{R}^d\rightarrow\mathbb{R}$ that is $\pi_k-$integrable, one is interested in estimating expectations w.r.t. the filtering distribution at observation times $(t_k)_{k\geq 1}$. These expectations are given by
\begin{equation*}
\pi_k(\varphi):= \int_{\mathbb{R}^d}\varphi(\mathbf{z}_{t_k}) \pi_k(\mathbf{z}_{t_k}) d\mathbf{z}_{t_k}.
\end{equation*}
After obtaining $N$ samples from $\pi_k^N$ in \eqref{eq:filter_approx}, the expectation $\pi_k(\varphi)$ can be estimated through
\begin{equation*}
\widehat{\pi}_k^N(\varphi) := \frac{1}{N} \sum_{i=1}^N \varphi(\mathbf{Z}_{t_k}^{(i)})
\end{equation*}
which as $N\to \infty$ converges to the true expectation $\pi_k(\varphi)$ almost surely \cite{smcmc}; a more precise statement is outlined in the Appendix.

\subsection*{Cost of SMCMC Algorithm}
The SMCMC method proved particularly effective and efficient in high dimensional problems and is of cost $\mathcal{O}([\kappa(d)+\kappa_y(d_y^k)]N)$ per update step where $\kappa(d)$ and $\kappa_y(d_y^k)$ are the costs for computing $f_k$ and $g_k$ at time $t_k$, respectively. The function $\kappa$ can be of the form $d$, $d\log d$ or $d^2$ depending on the structure of the state noise covariance matrix (similarly $\kappa_y$). In particular, the main cost comes from the vector-matrix multiplications included in evaluating the densities $f_k$ and $g_k$. If both matrices $Q_k$ (hidden signal noise covariance matrix at time $t_k$) and $R_k$ (data noise covariance matrix at time $t_k$) are diagonal, the cost of the SMCMC algorithm is $\mathcal{O}([d + d_y^k]N)$. If $Q_k$ is dense and $R_k$ is diagonal, the cost is of order $\mathcal{O}([d^2 + d_y^k]N)$. If $R_k$ is dense and $Q_k$ is diagonal, the cost is $\mathcal{O}([d + (d_y^k)^2]N)$. Finally, if both $Q_k$ and $R_k$ are dense, the cost is $\mathcal{O}([d^2+ (d_y^k)^2]N)$. On a multi-core machine, the vector-matrix operation gets distributed over the number of available cores thanks to libraries like BLAS and OpenBLAS.


In light of \eqref{eq:signal_unobserved}--\eqref{eq:obs_model}, the SMCMC filtering method of \cite{smcmc} is summarized in \autoref{alg:smcmc} in the Appendix. In practice, one would run $M$ independent runs of \autoref{alg:smcmc} in parallel then use averages for performing inference.

\section{Localized Sequential MCMC Filter}
\label{sec:lsmcmc}
In many practical applications, the observational locations tend to be or very sparse or highly localized in space. As an example of such applications is the set of data obtained from the SWOT mission led by NASA. It aims to provide high-resolution, detailed measurements across large swaths of oceans, lakes, and rivers with unprecedented accuracy collected via innovative interferometric radars. This set of data is expected to enhance predictive DA models by providing accurate initial and boundary conditions as well as accurate observations, thus improving forecasts for climate, weather, and hydrological systems. Another example, is the set of ocean data collected via the so-called ocean drifters, which are floating devices equipped with sensors that collect vital data on ocean currents, water velocities, and temperatures as they drift with the flow of the water. These drifters are deployed across the globe, but their distribution is often sparse, especially in remote regions like the polar seas or vast stretches of the open ocean. 

\begin{figure}[h!]
\centering
\includegraphics[scale=0.25, trim={2.2cm 4cm 3.5cm 4.5cm},clip]{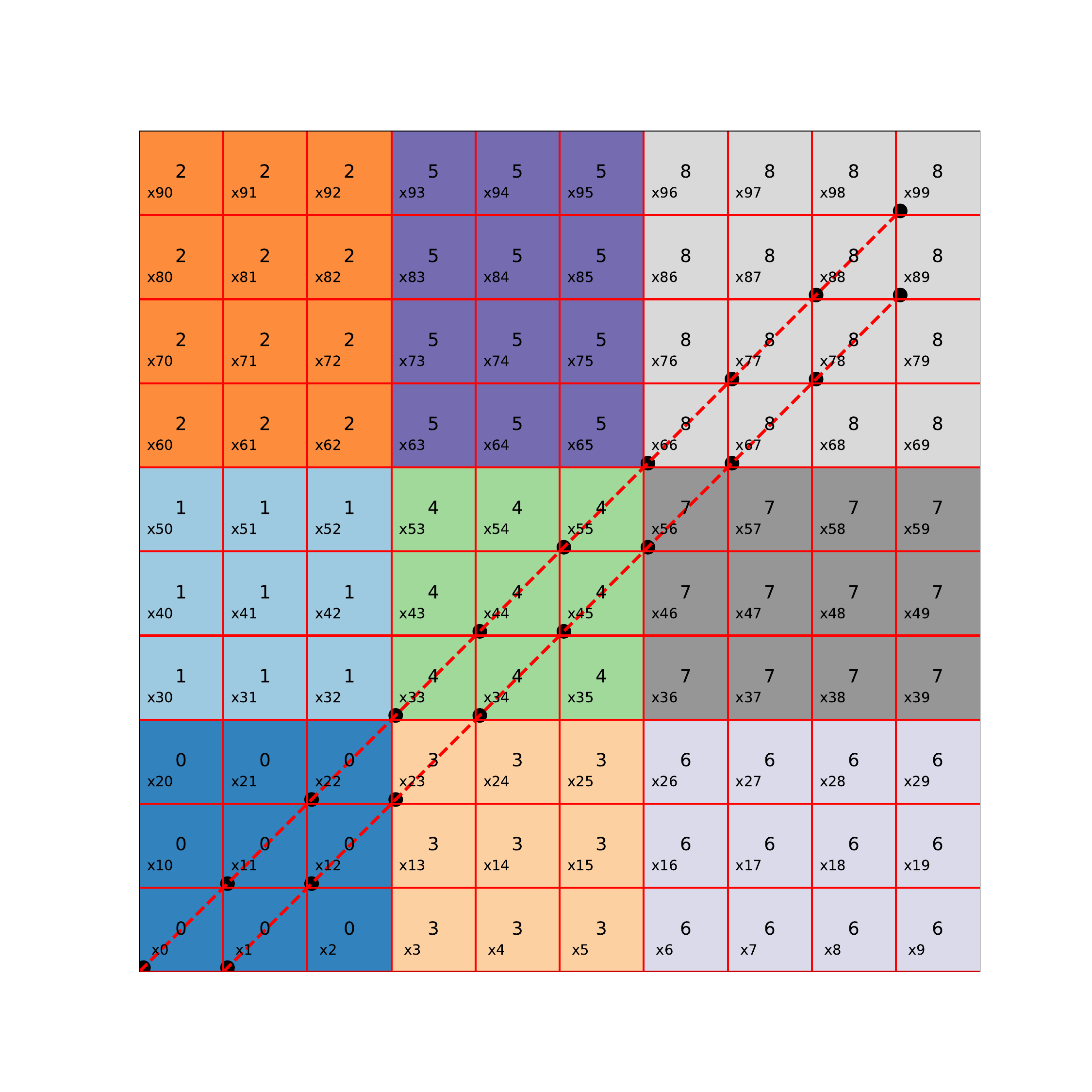}
\caption{Schematic showing a $10 \times 10$ square grid with nine subdomains labeled from 0 to 8 with a convention that the edges of a subdomain that contain points are the west and south edges. The black dots represent the locations of the observations. The subdomains labeled 0, 3, 4, 7 and 8 are the ones that contain observational locations and contain 55 grid points saved in $\overline{\mathbf{x}}_{t_k}$. Note that subdomains 3 and 7 only contain one observational location each, namely, $x_{23}$ and $x_{56}$, respectively.}
\label{fig:partition}
\end{figure}

In these situations, filtering the entire signal $\mathbf{Z}_{t_k}$ all at once would be computationally impractical. Instead, a more attractive approach would be to focus the filtering on the specific regions and time points where observations are available, rather than attempting to filter the entire signal as a whole. 
The basic idea behind our localization scheme is to perform a domain localization where the domain $\mathsf{G}$ is partitioned into $\Gamma$ subdomains $\mathsf{G}=\bigcup_{i=1}^{\Gamma} G_i$, and then, to find all subdomains $G_i$'s that contain observational locations and update the state variables in these subdomains. Note that this partition can be time-dependent; in other words, one can increase or decrease the number of subdomains $\Gamma$ at time $t$ to adopt the distribution of observations coming at this particular time. Let $\overline{\mathbf{x}}_{t_k} \subset \mathsf{G}$ denote the collection of grid points inside the subdomains that contain observational locations at time $t_k$. Let $\mathbf{z}_{t_k}(\overline{\mathbf{x}}_{t_k})\in \mathbb{R}^{d_k}$, $d_k < d$, be a sub-vector of the whole hidden signal $\mathbf{z}_{t_k}$ with variables at locations $\overline{\mathbf{x}}_{t_k}$. The steps of the algorithm are as follows. At observational time $t_1$, one draws the samples $\{\mathbf{Z}_{t_1}^{(i)}(\overline{\mathbf{x}}_{t_1})\}_{1\leq i\leq N}$ from the distribution
\begin{equation}
\label{eq:loc_pi_1}
\widetilde{\pi}_1(\mathbf{z}_{t_1}(\overline{\mathbf{x}}_{t_1})) \propto \widetilde{g}_1(\mathbf{z}_{t_1}(\overline{\mathbf{x}}_{t_1}),\mathbf{y}_{t_1})~\widetilde{f}_1(\mathbf{z}_0(\overline{\mathbf{x}}_{t_1}),\mathbf{z}_{t_1}(\overline{\mathbf{x}}_{t_1})),
\end{equation}
and at time $t_k$, $k>1$, draws the samples $\{\mathbf{Z}_{t_k}^{(j_i)}(\overline{\mathbf{x}}_{t_k}),~ j_i\}_{1\leq i \leq N}$ from the joint distribution
\begin{equation}
\label{eq:filter_local}
\widetilde{\pi}_k^{N}(\mathbf{z}_{t_k}(\overline{\mathbf{x}}_{t_k}),j) \propto \widetilde{g}_k(\mathbf{z}_{t_k}(\overline{\mathbf{x}}_{t_k}),\mathbf{y}_{t_k}) ~\widetilde{f}_k(\mathbf{z}_{t_{k-1}}^{(j)}(\overline{\mathbf{x}}_{t_k}),\mathbf{z}_{t_k}(\overline{\mathbf{x}}_{t_k}))~p(j),
\end{equation}
using an appropriate MCMC kernel in both cases. The density $\widetilde{f}_k$ can be thought of as the restriction of the transition density $f_k$ on subdomains $G_i$'s with locations $\overline{\mathbf{x}}_{t_k}$, and $\widetilde{g}_k$ is the likelihood density resulting from a modified observational model $\widetilde{\mathscr{O}}_{t_k}:\mathbb{R}^{d_k}\to \mathbb{R}^{d_y^k}$. For instance, if $\mathscr{O}_{t_k}=C\mathbf{z}_{t_k}$ for some matrix $C\in \mathbb{R}^{d_y^k\times d}$, then one would modify $C$ by keeping only the columns that correspond to $\mathbf{z}_{t_k}(\overline{\mathbf{x}}_{t_k})$ and have $\widetilde{\mathscr{O}}_{t_k}=\widetilde{C}\mathbf{z}_{t_k}(\overline{\mathbf{x}}_{t_k})$, where $\widetilde{C}\in \mathbb{R}^{d_y^k\times d_k}$ is the new modified matrix. After obtaining the samples $\mathbf{z}_{t_k}(\overline{\mathbf{x}}_{t_k})$, one updates the rest of the variables (that is, update $\mathbf{z}_{t_k}(\overline{\mathbf{x}}_{t_k}^c)$, where $\overline{\mathbf{x}}_{t_k}^c$ refers to the complement $\mathsf{G}\setminus \overline{\mathbf{x}}_{t_k}$) according to the dynamics in the forward model. In light of \eqref{eq:signal_unobserved}--\eqref{eq:obs_model}, for a given $j\in \{1,\cdots,N\}$, one would have 
\begin{equation}
\widetilde{f}_k\left(\mathbf{Z}_{t_{k-1}}^{(j)}(\overline{\mathbf{x}}_{t_k}),\mathbf{z}_{t_k}(\overline{\mathbf{x}}_{t_k})\right) \propto \exp\left\{-\frac{1}{2}[\mathbf{z}_{t_k}(\overline{\mathbf{x}}_{t_k})-\mathbf{Z}_{t_{k-1}}^{(j)}(\overline{\mathbf{x}}_{t_k})]^\top\widetilde{Q_k^{-1}}[\mathbf{z}_{t_k}(\overline{\mathbf{x}}_{t_k})-\mathbf{Z}_{t_{k-1}}^{(j)}(\overline{\mathbf{x}}_{t_k})]\right\},
\end{equation}
where $\widetilde{Q_k^{-1}}$ is the matrix with rows and columns of $Q_k^{-1}$ that correspond to the grid subdomains that contain the locations $\overline{\mathbf{x}}_{t_k}$; see \autoref{fig:partition} for an illustration. The figure shows a $10 \times 10$ square grid partitioned into 9 subdomains and shows observations at 19 locations. In the case of having one hidden variable per grid point, this partitioning shown in \autoref{fig:partition} will result in a large dimension reduction by replacing the matrix $Q_k^{-1} \in \mathbb{R}^{100\times 100}$ with the matrix $\widetilde{Q_k^{-1}}\in \mathbb{R}^{55\times 55}$. The steps of the localization scheme are summarized in \autoref{alg:loc_smcmc}. Similar to the SMCMC algorithm, one would run \autoref{alg:loc_smcmc} $M$ times in parallel then average over the multiple runs. We note that the localization described here is not limited to SSMs with Gaussian noise, but it can be applied to any SSM as long as the noise covariance matrix $Q_k$ and its inverse can be computed or estimated. 

%
%
\begin{flushleft}
\needspace{4\baselineskip}
\captionsetup[algorithm]{style=algori}
\captionof{algorithm}{The Scheme of LSMCMC}
\label{alg:loc_smcmc}
\textbf{Input:} Given the matrices $Q_k$ and $Q_k^{-1}$, a rectangular partition of the grid $\mathsf{G}=\bigcup_{i=1}^{\Gamma} G_i$, the initial state $\breve{\mathbf{Z}}_0=\mathbf{z}_0$, the observations $\{\mathbf{Y}_{t_k} = \mathbf{y}_{t_k}\}_{k\geq 1}$ and their locations, and the number of time discretization steps (observational time-lag) $L$. Set $\tau_k:=(t_k-t_{k-1})/L$.

\begin{enumerate}
\item Initialize: 
\begin{enumerate}
\item For $l=0,\cdots,L-1$ compute $\breve{\mathbf{Z}}_{(l+1)\tau_1} = \Phi(\breve{\mathbf{Z}}_{l\tau_1},l\tau_1;(l+1)\tau_1)$.
\item Find the subdomains that contain observational locations at time $t_1$ and return the unique grid points $\overline{\mathbf{x}}_{t_1}$ and the sub-matrix $\widetilde{Q_1^{-1}}$.
\item Compute $\tilde{\mathbf{Z}}_{t_1}:=\breve{\mathbf{Z}}_{t_1}+ \mathbf{W}_{t_1}$, where $\mathbf{W}_{t_1}\sim \mathcal{N}_d(0,Q_1)$.
\item Run a standard RWM initialised at $\tilde{\mathbf{Z}}_{t_1}(\overline{\mathbf{x}}_{t_1})$ to generate $N$ samples $\{\mathbf{Z}_{t_1}^{(i)}(\overline{\mathbf{x}}_{t_1})\}_{i=1}^N$ from $\widetilde{\pi}^N_1$ in \eqref{eq:loc_pi_1}. For $i=1,\cdots,N$, set $\mathbf{Z}_{t_1}^{(i)}(\overline{\mathbf{x}}_{t_1}^c) \longleftarrow \tilde{\mathbf{Z}}_{t_1}(\overline{\mathbf{x}}_{t_1}^c)$.
\item Set $\widehat{\pi}_{1}^N(\varphi) \leftarrow \frac{1}{N} \sum_{i=1}^N \varphi(\mathbf{Z}_{t_1}^{(i)})$.
\end{enumerate}

\item For $k=2,\ldots,T$: 
\begin{enumerate}
\item For $i=1,\cdots, N$: Compute 
$$\breve{\mathbf{Z}}_{(l+1)\tau_k+t_{k-1}}^{(i)} = \Phi(\breve{\mathbf{Z}}_{l\tau_k+t_{k-1}}^{(i)},l\tau_k+t_{k-1};(l+1)\tau_k+t_{k-1}),$$ 
where $0\leq l \leq L-1$ and $\breve{\mathbf{Z}}^{(i)}_{t_{k-1}}= \mathbf{Z}^{(i)}_{t_{k-1}}$. 
\item Find the subdomains that contain observational locations at time $t_1$ and return the unique grid points $\overline{\mathbf{x}}_{t_k}$ and the sub-matrix $\widetilde{Q_k^{-1}}$. 
\item For $i=1,\cdots, N$: Compute $\tilde{\mathbf{Z}}_{t_k}^{(i)}:=\breve{\mathbf{Z}}_{t_k}^{(i)}+ \mathbf{W}_{t_k}^{(i)}$, where $\mathbf{W}_{t_k}^{(i)}\sim \mathcal{N}_d(0,Q_k)$.
\item Run \autoref{alg:loc_rwm} to return the samples $\{\mathbf{Z}_{t_k}^{(i)}(\overline{\mathbf{x}}_{t_k})\}_{i=1}^N$ from \eqref{eq:filter_local}. For $i=1,\cdots,N$, set $\mathbf{Z}_{t_k}^{(i)}(\overline{\mathbf{x}}_{t_k}^c) \longleftarrow \tilde{\mathbf{Z}}_{t_k}^{(i)}(\overline{\mathbf{x}}_{t_1}^c)$.
\item Set $\widehat{\pi}_{k}^N(\varphi) \leftarrow \frac{1}{N} \sum_{i=1}^N \varphi(\mathbf{Z}_{t_k}^{(i)})$. 
\end{enumerate}
\end{enumerate}

\textbf{Output:} Return $\{\widehat{\pi}_k^N(\varphi)\}_{k\in\{1,\cdots,T\}}$.

\vspace{-0.1cm}
\hrulefill
\vspace{0.2cm}
\end{flushleft}

%
%
\begin{flushleft}
\captionsetup[algorithm]{style=algori}
\captionof{algorithm}{Pseudocode for RWM to sample from $\widetilde{\pi}_k^{N}(z_{t_k}(\overline{\mathbf{x}}_{t_k}),j)$}
\label{alg:loc_rwm}

\textbf{Initialization:} 
\begin{itemize}
\item Sample the auxiliary variable $j_0 \sim p(j)$ (uniformly).
\item Set $\mathbf{Z}_{t_k}^{(0)} \longleftarrow \tilde{\mathbf{Z}}^{(j_0)}_{t_k}$. 
\item Compute
$
\pi_{\text{old}} \longleftarrow  \widetilde{g}_k(\mathbf{Z}_{t_k}^{(0)}(\overline{\mathbf{x}}_{t_k}),\mathbf{y}_{t_k}) ~\widetilde{f}_k(\mathbf{Z}_{t_{k-1}}^{(j_0)}(\overline{\mathbf{x}}_{t_k}),\mathbf{Z}_{t_k}^{(0)}(\overline{\mathbf{x}}_{t_k})).$ 
\end{itemize}
For $i ={1,\ldots,N+N_{\text{burn}}}$:
\begin{enumerate}

\item Compute proposal for the:
\begin{itemize}
\item  state: $\mathbf{Z}'_{t_k} \longleftarrow \mathbf{Z}_{t_k}^{(i-1)} + \mathbf{W}'$, where $\mathbf{W}' \sim \mathcal{N}_d(0,{Q}')$ for some covariance matrix $Q'\in \mathbb{R}^{d\times d}$; 
\item auxiliary variable: $j' = \left\{\begin{array}{ll}
		j_{i-1}-1 & \textrm{if}~j_{i-1}\notin\{1,N\}~\textrm{{w.p.}}~q\\
		j_{i-1} &  \textrm{if}~j_{i-1}\notin\{1,N\}~\textrm{{w.p.}}~1-2q\\
		j_{i-1}+1 & \textrm{if}~j_{i-1}\notin\{1,N\}~ \textrm{{w.p.}}~q\\
j_{i-1}+1 & \textrm{if}~j_{i-1}=1\\
j_{i-1}-1 & \textrm{if}~j_{i-1}=N
		\end{array}\right.$, where $q\in(0,\frac{1}{2}]$.
\end{itemize}
\item Compute 
$
\pi_{\text{new}} \longleftarrow  \left\{\begin{array}{ll}
		\widetilde{g}_k(\mathbf{Z}_{t_k}'(\overline{\mathbf{x}}_{t_k}),\mathbf{y}_{t_k}) ~\widetilde{f}_k(\mathbf{Z}_{t_{k-1}}^{(j')}(\overline{\mathbf{x}}_{t_k}),\mathbf{Z}_{t_k}'(\overline{\mathbf{x}}_{t_k})) & \textrm{if}~j_{i-1}\notin\{1,N\}\\
		\widetilde{g}_k(\mathbf{Z}_{t_k}'(\overline{\mathbf{x}}_{t_k}),\mathbf{y}_{t_k}) ~\widetilde{f}_k(\mathbf{Z}_{t_{k-1}}^{(j')}(\overline{\mathbf{x}}_{t_k}),\mathbf{Z}_{t_k}'(\overline{\mathbf{x}}_{t_k})) ~q &  \textrm{if}~j_{i-1}\in\{1,N\}
		\end{array}\right.  .$ 
\item  Compute $\alpha = \min\{1, \pi_{\text{new}}/\pi_{\text{old}}\}$. Sample $u \sim \mathcal{U}[0,1]$. \begin{itemize}
\item  If $ u<\alpha$ set $\mathbf{Z}_{t_k}^{(i)}\longleftarrow \mathbf{Z}_{t_k}'$, $j_i\longleftarrow j'$ and $\pi_{\text{old}}\longleftarrow\pi_{\text{new}}$.
\item Else set $\mathbf{Z}_{t_k}^{(i)}\longleftarrow \mathbf{Z}_{t_k}^{(i-1)}$ and $j_{i} \longleftarrow j_{i-1}.$
\end{itemize}
\end{enumerate}

\textbf{Output:} Return the sequence of $N$ final samples $\left\{\mathbf{Z}_{t_k}^{(i)}(\overline{\mathbf{x}}_{t_k})\right\}_{i=N_{\text{burn}}+1}^N$.

\vspace{-0.1cm}
\hrulefill
\vspace{0.2cm}
\end{flushleft}

In many applications, the covariance matrix is often time-independent, and therefore, its inverse or the inverse of its Cholesky factor is computed only once at the beginning. In this case, we drop the subscript $k$ in $Q_k$ and $Q_k^{-1}$ in the above algorithm.

\section{Numerical Simulations}
\label{sec:numerics}
We will illustrate the performance of LSMCMC filter (\autoref{alg:loc_smcmc}) in three challenging cases:
\begin{enumerate}
    \item A linear Gaussian model on a 2d grid that is partially observed in space. At each observation time $t_k$, the data collected forms a swath of a certain width. This width is relatively small compared to the overall width of the grid. The swath is oriented at an angle, either tilting to the right or to the left, and this direction changes at alternate observation times. This swath of data moves from east to west in a cyclical manner. To highlight the advantages of the proposed algorithm, we compare the results obtained using \autoref{alg:loc_smcmc} with the original SMCMC \autoref{alg:smcmc}, the EnKF algorithm and the LEnKF \autoref{alg:LEnKF}.
    \item A SSM with nonlinear rotating shallow water equations (RSWEs) observed in a similar fashion as in the previous case in a domain of the Atlantic Ocean. In this case, the swath of observations is for the water heights where we try to mimic the SWOT data scenario. The simulation initial and boundary conditions are obtained from Copernicus Marine Service \cite{coper} to make the case study as realistic as possible. We show that for a given accuracy and the same number of samples, LSMCMC is about 5.6 times faster than SMCMC.
    \item A SSM with RSWEs observed with ocean drifters of uncertain observational locations. This example is a replicate of the fourth model in \cite{smcmc}. We use the same real data used in that paper for the drifter positions and velocities which are obtained from NOAA; see \cite{noaa, adam1}. We show that the proposed localization algorithm yields similar results as the original SMCMC filter, but it is significantly faster.
\end{enumerate}

In the first example, the model is tractable where the filter mean is available through the Kalman filter (KF), and therefore, it is easy to compare the different algorithms. The second and third examples are high-dimensional nonlinear models which are non-tractable. In the second example, although the initial and boundary conditions are realistic, the observations are synthetic, i.e., they are generated from one run of the forward model. Hence, we compare the filter mean to the result of this particular run, which is a common metric used in the literature when exact filter mean is unavailable. In the third example, the observational data is realistic. In this case, we compare our result against an average of 50 runs of the forward model with the same initial and boundary conditions.

All simulations were run on a Precision 7920 Tower Workstation with 52 cores and 512GB of memory. The EnKF and LEnKF algorithms implemented in the first example are run once, therefore, making full use of all available cores when performing matrix-matrix multiplications or solving linear systems. As for the SMCMC and LSMCMC algorithms in the second and third examples, we implement 26 independent runs allowing for multi-threading on the 26 cores left. 

The partitioning algorithm we use results in subdomains that are square or rectangular with sides that are close to each other in length. To achieve this, the following condition must be met. The parameter $\Gamma$ should divide $(N_g^x-1)(N_g^y-1)$ and the result of this division should not be a prime integer. If this condition is not met, the partitioning algorithm looks for a another $\Gamma$ that satisfies this requirement.

\subsection{Linear Gaussian Model with SWOT-Like Observations}
\label{subsec:linear_model}
\begin{figure}[h!]
\centering
\includegraphics[scale=0.17]{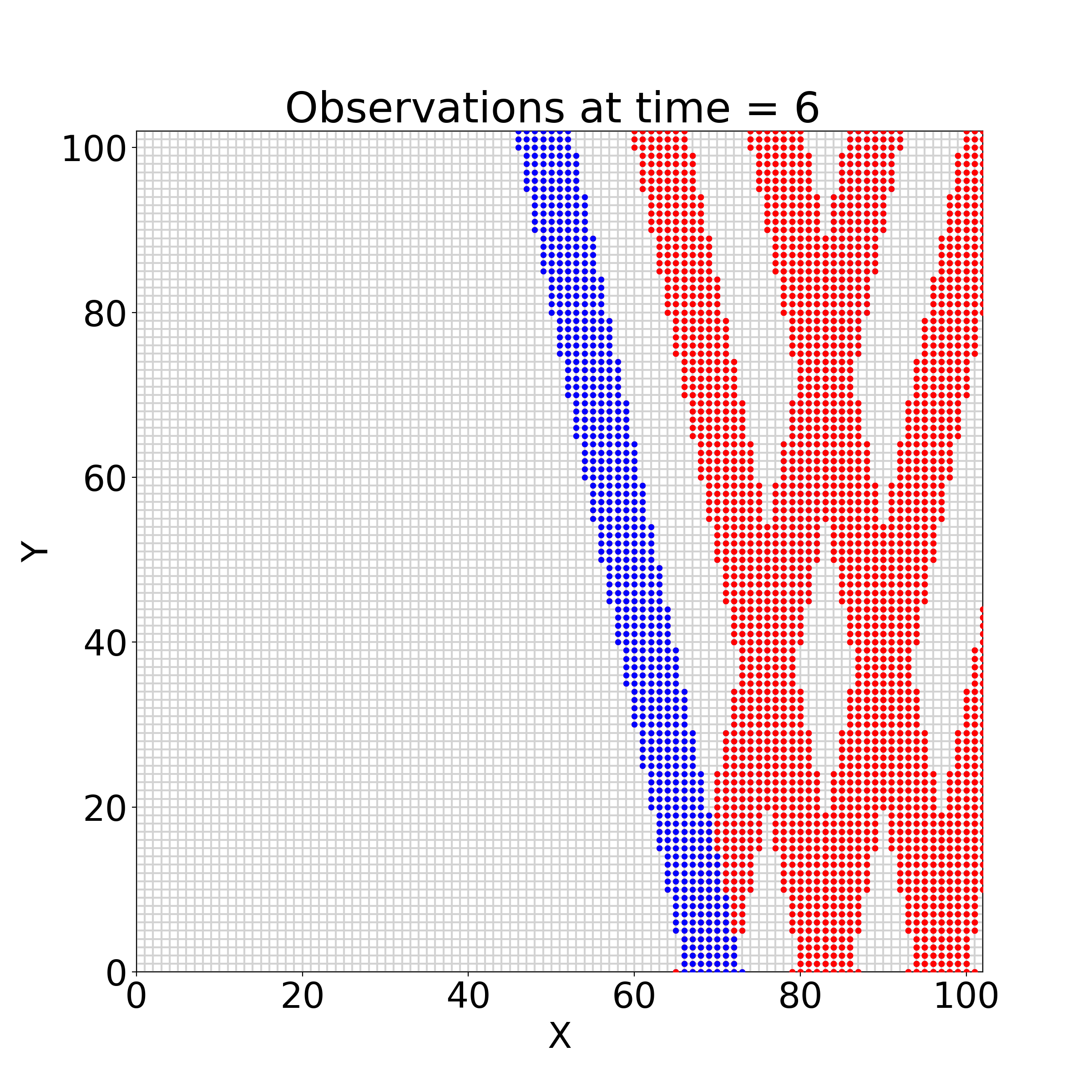}
\includegraphics[scale=0.17]{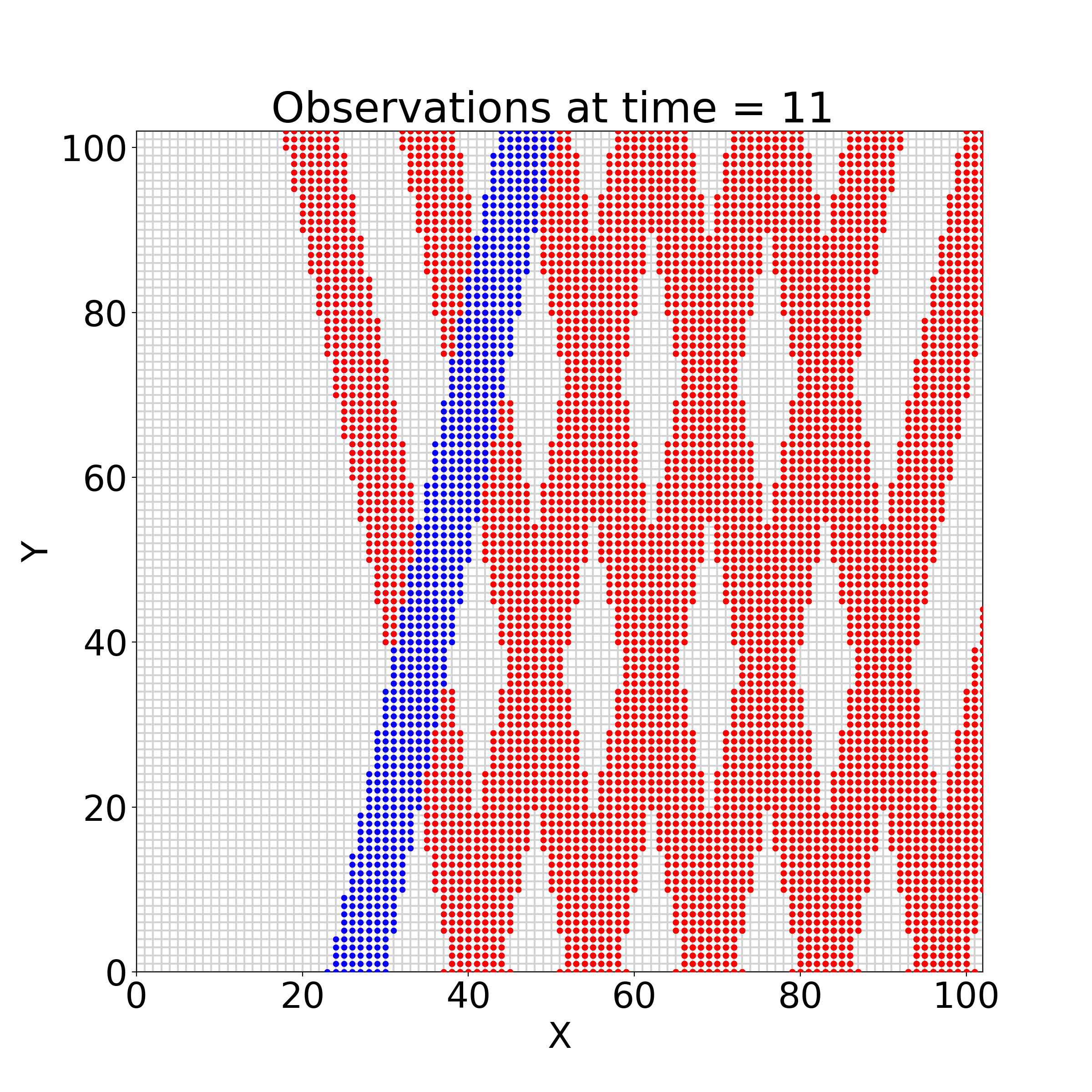}
\includegraphics[scale=0.17]{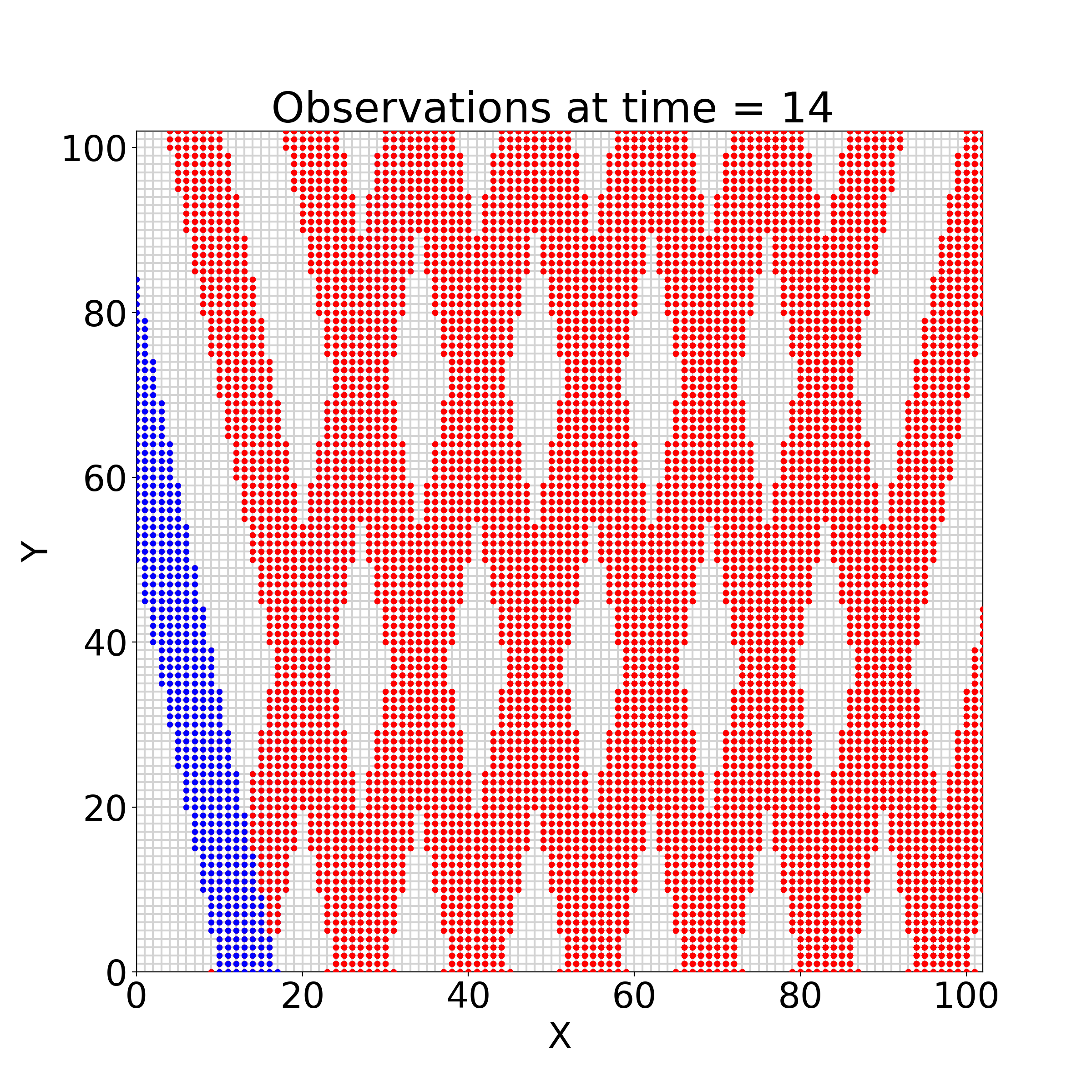}
\caption{These images show the swaths of observations at different times for the linear Gaussian model. The shots show the observations at times 6, 11 and 14 (in blue), respectively, and the swaths of the observations at previous times (in red). In the assimilation step, only the blue swath of data is used.}
\label{fig:linear_obs_swaths}
\end{figure}

Consider the forward model
\begin{equation}
\mathbf{Z}_{k+1} = A \mathbf{Z}_k + \sigma_z \mathbf{W}_{k+1}, \quad \mathbf{W}_{k+1} \stackrel{\textrm{i.i.d.}}{\sim} \mathcal{N}_d(0,I_d),  \quad k\in \{0,\cdots,T\},
 \label{eq:linear_model}
\end{equation}
where $T\in \mathbb{N}$, $\mathbf{Z}_{k+1}\in \mathbb{R}^{d}$ with $d=N_g^x \times N_g^y$, $A\in \mathbb{R}^{d \times d}$ is a square matrix where its maximum eigenvalue is less than or equal to one. For $k\in \{1,\cdots,T\}$, the observations are generated from a single run of the forward model. The observations are a swath of data of a width of 7 grid points inclined with some angle $\theta$ that moves from east to west in a cyclic way. If at time $k$ the angle $\theta <\pi/2$, then at time $k+1$ we have $\theta > \pi/2$; see \autoref{fig:linear_obs_swaths}. The dimension of the data $d_y^k=m_k$, $m_k<N_g^x N_g^y$, might change from time to time. For a given time $k\in \{1,\cdots,T\}$, the observations $\mathbf{Y}_k \in \mathbb{R}^{m_k}$ are at locations (grid points) $\mathbf{x}_k=\{x_k^{(1)}, \cdots, x_k^{(m_k)}\}$ defined as $x_k^{(n)} := i_k^n+(j_k^n-1)N_g^x$, where $\{i_k^n\}_{n=1}^{m_k} \subset \{1,\cdots, N_g^x\}$ and $\{j_k^n\}_{n=1}^{m_k} \subset \{1,\cdots, N_g^y\}$. The observations at time $k$ follow
\begin{equation}
\label{eq:numer_obs_model}
\mathbf{Y}_k = C\mathbf{Z}_k +\sigma_y \mathbf{V}_k, \quad \mathbf{V}_k \stackrel{\textrm{i.i.d.}}{\sim} \mathcal{N}_d(0,I_{d_y^k}).
\end{equation}
The matrix $C\in \mathbb{R}^{d_y^k \times d}$ is defined as $C=[C_{m,n}]$ with elements 
\begin{align}
\label{eq:matrix_C}
C_{m,n} = \left\{\begin{array}{ll}
1 & \text{if } n = x_k^{(m)}\\
0 & \text{otherwise}
\end{array} \right..
\end{align}

We compare the LSMCMC with the SMCMC scheme described in \autoref{alg:smcmc}. We also compare it with the EnKF in which we use the matrix inversion lemma (Sherman-Morrison-Woodbury formula) when $\max_{k\in \{1,\cdots, T\}} d_y^k$ is larger than the ensemble size; see \cite{nino,mandel,vetra} for the pseudocodes and more details. We should note that when we implement the EnKF or its localized version (which will be described shortly) we \emph{do not} directly compute the inverse of a matrix if it is being multiplied from left or right by another matrix or vector. Rather we reformulate the problem as finding the solution of a linear system. Finally, we also compare the LSMCMC with an LEnKF scheme \cite{hunt, vetra} which is an explicit localization technique referred to as the $R$-localization. The basic idea is to partition the domain into $\Gamma$ subdomains as before, then, to only use the observations within a specified distance around every local subdomain $G_i$. This defines a linear transformation, which transforms the observation error covariance matrix $R_k$, the global observation vector $\mathbf{Y}_k$, and the global observation matrix $C$ to their local parts; see \cite{vetra} and the pseudocode in the Appendix. Then, an observation localization is performed by modifying the observation error covariance matrix $R_k$ so that the inverse observation variance decreases to zero with the distance of an observation from an analysis grid point. The modification is done by dividing the diagonal elements of $R_k$ by some weights that are a function (e.g. Gaspari--Cohn function \cite{gaspari}) of the distance of observations from the analysis grid point scaled by a positive number $r$ measured in grid points. The updates on the local domains can be done independently and therefore we do them in parallel \cite{hunt}. A pseudocode for the LEnKF used in this paper is included in the Appendix in \autoref{alg:LEnKF}. 

We use two metrics here. The first metric is to measure the percentage of absolute errors that are less than $\sigma_y/2$. The absolute errors are saved in a matrix of size $T\times d$ and defined as the absolute difference of the mean of the KF and the mean of the filter of interest at every time step $k$ and every state variable $Z_k^{(j)}$, $j\in \{1,\cdots,d\}$. The intuition behind this is to measure how far is the filter mean from the observations. The second metric is to measure the computational time needed for each filter. Furthermore, we perform three tests for this high-dimensional linear Gaussian example. 

\subsection*{Simulation Settings and Results}

We set $T=100$, $N_g^x=N_g^y= 103$ and therefore $d=10609$, $\sigma_z=\sigma_y=0.05$, $A=0.25~I_d$ and $Z_0^{(j)}\sim -0.15 \times \mathcal{U}_{[0,1]}$ for all $j\in \{1,\cdots,\lfloor d/3 \rfloor\}$ and $Z_0^{(j)} = 0$ for the rest of $j$'s. In both SMCMC and LSMCMC we set the number of independent runs $M$ to 50.

\begin{figure}[h!]
\centering
\includegraphics[scale=0.2]{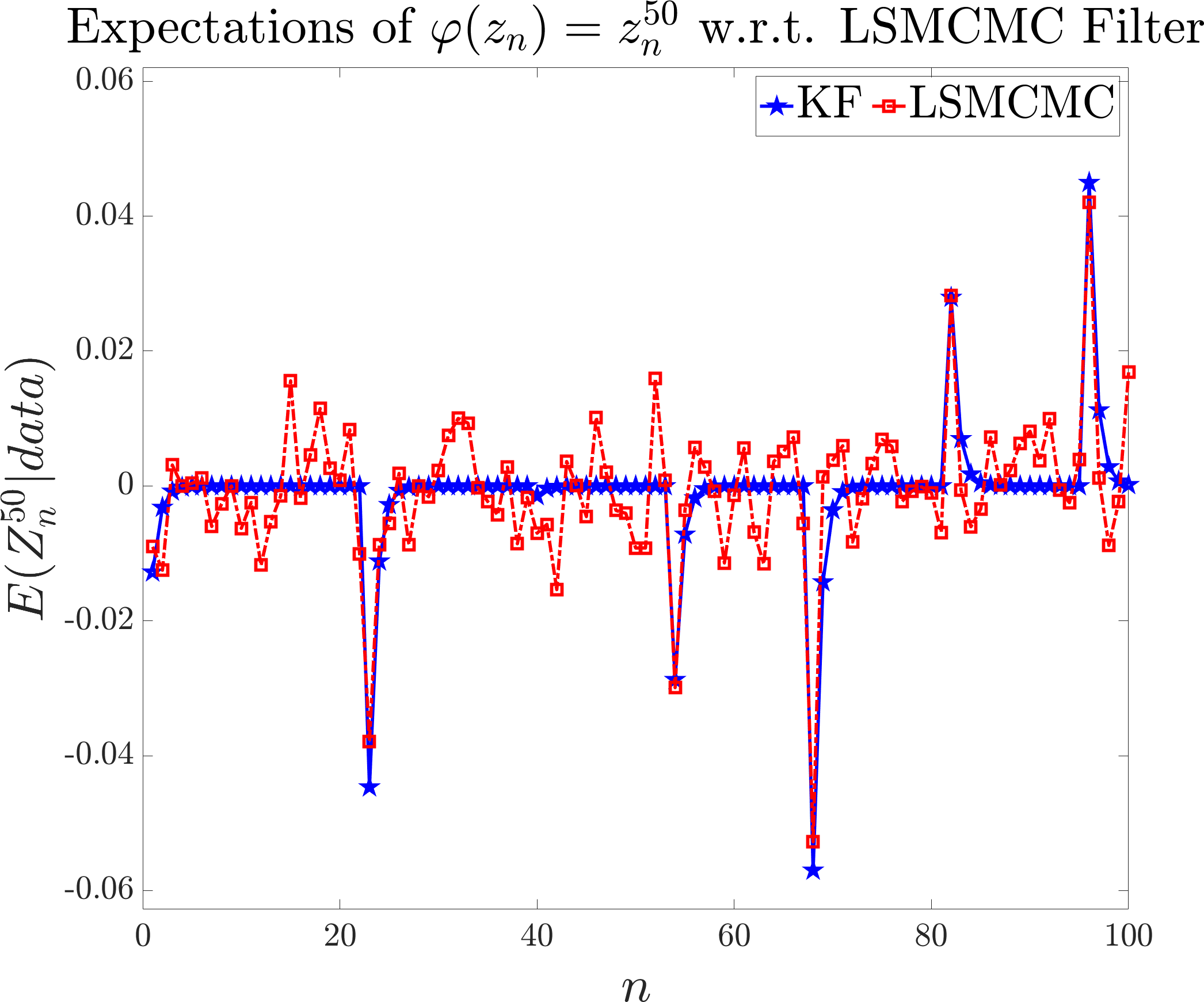}\,
\includegraphics[scale=0.2]{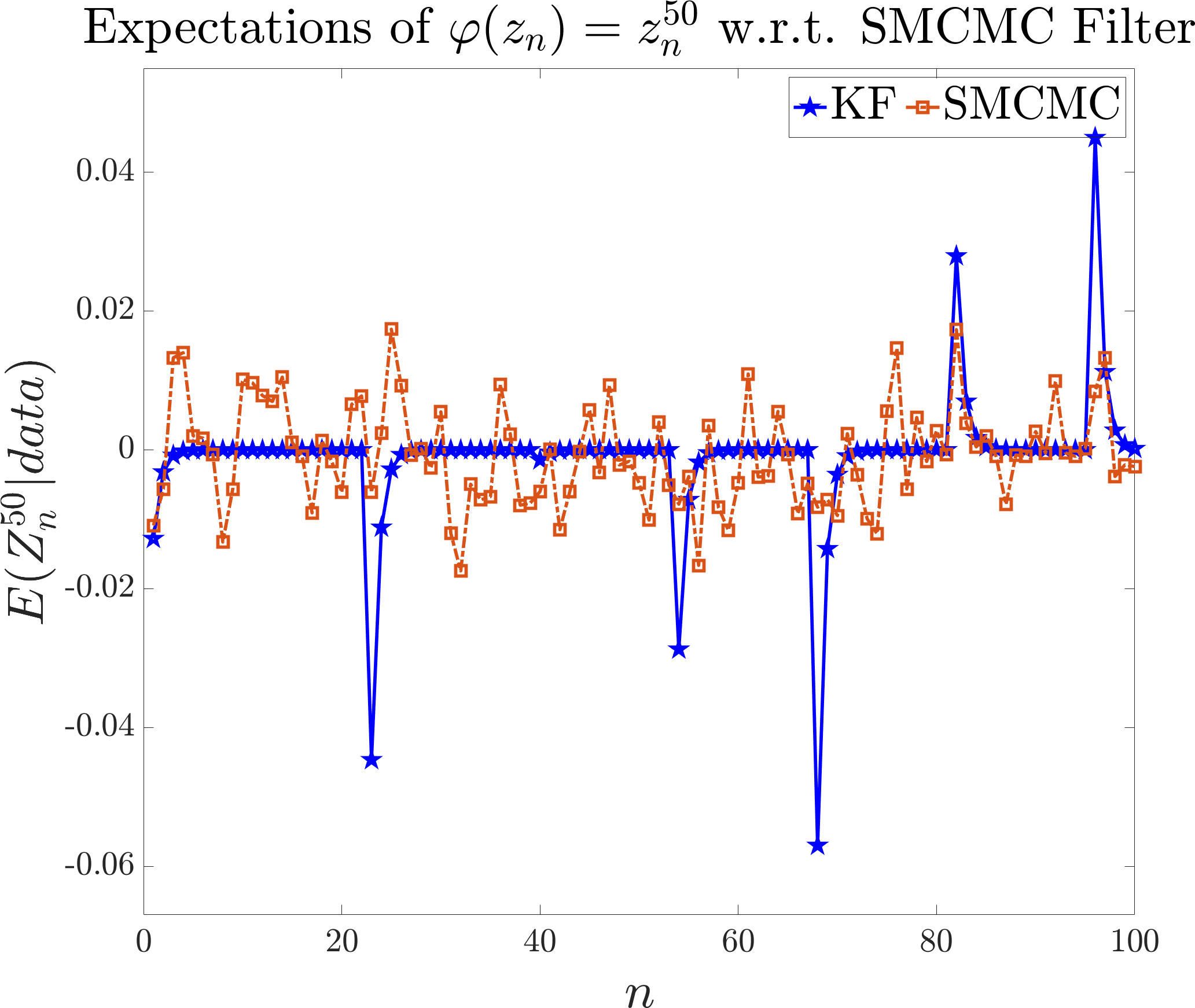}\\
\smallskip
\includegraphics[scale=0.2]{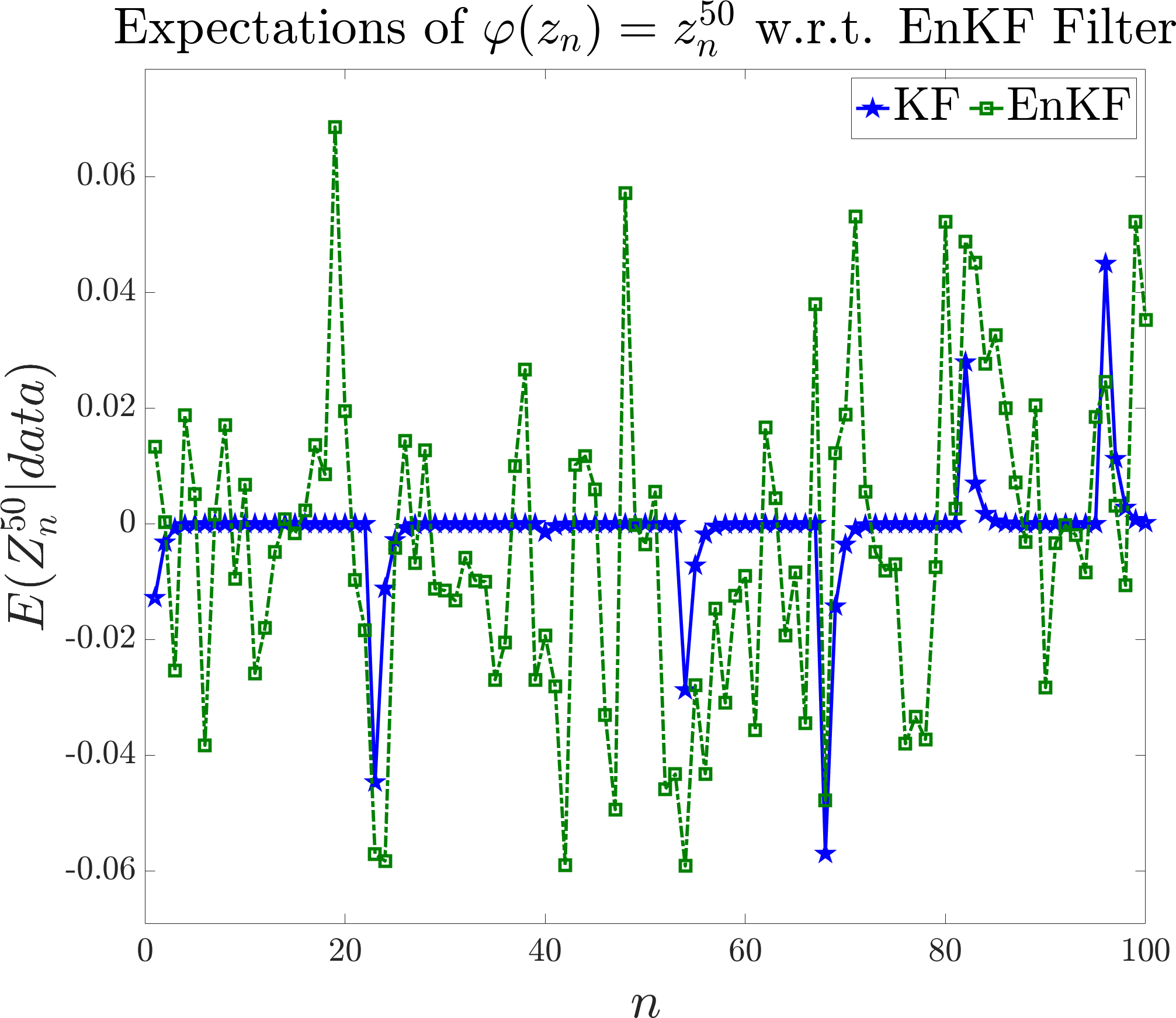}\,
\includegraphics[scale=0.2]{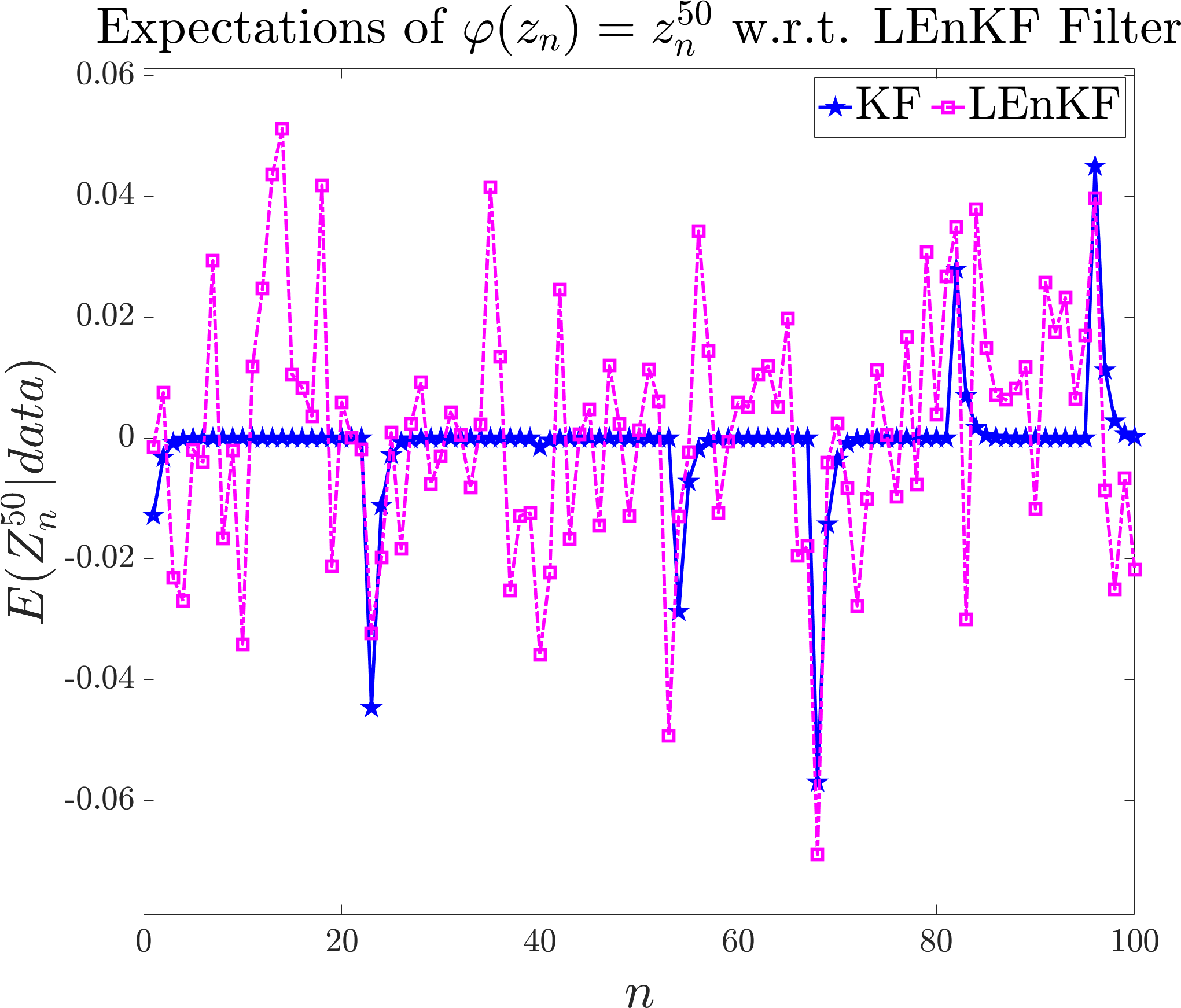}
\caption{Snapshots that show the filter mean at coordinate 50 of the different filtering schemes for the linear Gaussian model. The blue line is the true filter mean computed via KF.}
\label{fig:linear_coord50}
\end{figure}

\begin{figure}[h!]
\centering
\includegraphics[scale=0.5]{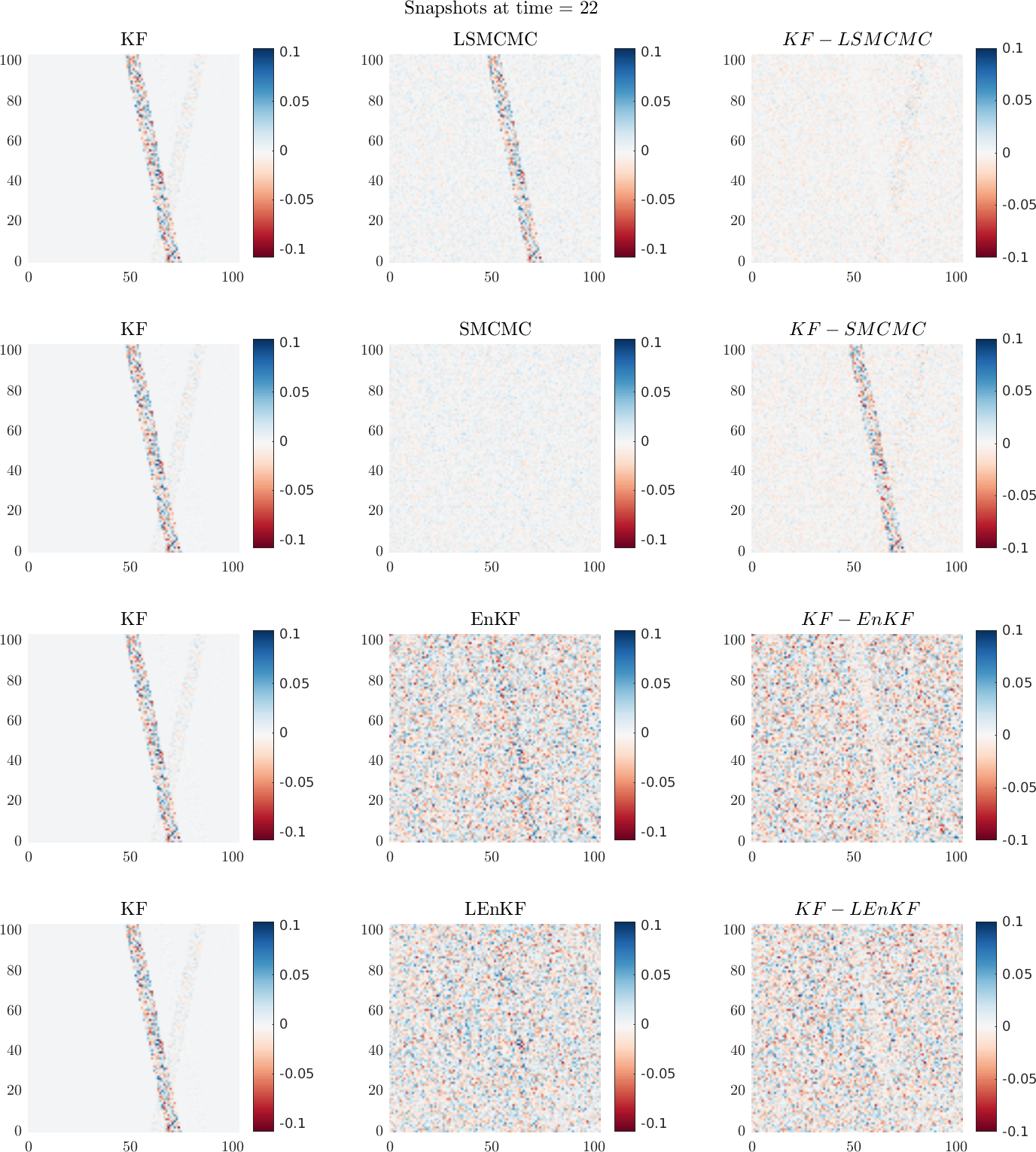}
\caption{Snapshot of the different filters means for the linear Gaussian model at time 22 using the same parameters as in \autoref{tab:table1}. The KF is on the very left, the second column is the filters means and the last column on the right is differences. }
\label{fig:linear_lsmcmc_vs_others}
\end{figure}

\textbf{First test:} We choose $\Gamma$, $N$ (and $N_{\text{burn}}$),  $r$ and $M$ so that all methods take the same computational time, which is set to 145 seconds. This is the time needed so that at least one of the methods results in at least 70\% of the absolute errors to be less than $\sigma_y/2$. The results of the first test are presented in \autoref{tab:table1} and  \Crefrange{fig:linear_coord50}{fig:linear_lsmcmc_vs_others}. The large dimension reduction that results from the localization of the SMCMC scheme allows us to increase the number of MCMC samples (and the burn-in samples) while maintaining a lower cost as shown in the first two rows of \autoref{tab:table1}. As we can see from the table, the LSMCMC results in 99.79\% of absolute errors being less than $\sigma_y/2$ versus 97.15\% of SMCMC, 70.74\% of EnKF and 73.27\% of LEnKF for the same computational cost. The reason for the relative low accuracy of EnKF and LEnKF is that they fail to accurately predict the unobserved variables of $\mathbf{Z}_k$ as shown in \autoref{fig:linear_coord50}. There, we plot the means of the filters against the mean of the KF at a particular variable of $\mathbf{Z}_k$, which we choose to be the 50$^{th}$ variable. As can be seen, the KF is almost zero everywhere in the time window except at times where the observations swath passes by the location $x_k^{(50)}$. We can see that the LSMCMC filter almost coincides with the KF at these times and oscillates around the curve of the KF elsewhere, but the oscillations are within a small interval $[-0.016, 0.018]$. The SMCMC algorithm fails to coincide with the KF at these times (except probably at the first and the one before the last) because of the low number of MCMC samples used; however, the oscillations around the KF curve elsewhere are also within a small interval $[-0.017, -0.018]$. As for the EnKF and LEnKF they almost coincide with the KF at two and four time instants, respectively, but their oscillations about the KF curve elsewhere are much wider on intervals $[-0.06,0.07]$ and $[-0.073, 0.052]$, respectively. In \autoref{fig:linear_lsmcmc_vs_others}, we show a snapshot of the different filters on the whole grid at assimilation time 22 using the parameters in \autoref{tab:table1}. In the snapshot we can see that the errors are lowest for LSMCMC. 

\begin{table}[h!]
\begin{tabular}{|c|c|c|c|c|c|c|}
\hline
\backslashbox{Schemes}{Parameters} &
  $\Gamma$ &
  $r$ &
  $M$ &
  \begin{tabular}[c]{@{}c@{}}$N$ \\ (or $N+N_{\text{burn}}$)\end{tabular} &
  \begin{tabular}[c]{@{}c@{}}Computational\\ time (seconds)\end{tabular} &
  \begin{tabular}[c]{@{}c@{}}\% of absolute\\ errors $<\sigma_y/2$\end{tabular} \\ \hline
LSMCMC & 1156 & -- & 52  & $5000+3000$ & 145.51 & 99.79\% \\ \hline
SMCMC  & --   & -- & 52 & $1000+700$  & 145.72 & 97.15\% \\ \hline
EnKF   & --   & -- & 1  & 1200        & 145.64 & 70.74\% \\ \hline
LEnKF  & 36  & 130 & 1 & 950         & 144.91 & 73.27\% \\ \hline
\end{tabular}
\caption{The results of the first test on the linear Gaussian model. The parameters are chosen such that the computational time is almost the same and that the percentage of absolute errors which are less than $\sigma_y/2$ is larger than 70\%.}
\label{tab:table1}
\end{table}

\textbf{Second test:} We set the number of MCMC and burn-in samples in SMCMC and the ensemble size in EnKF and LEnKF to be the same as in LSMCMC in the previous test (i.e. $N=5000$) and record the percentage of absolute errors less than $\sigma_y/2$ and the computational time in \autoref{tab:table2}. When $N$ is the same for all schemes, the LSMCMC is dominating in terms of accuracy and computational time. For example, when $\Gamma=36$ and $r=130$, it took LEnKF 545.13 seconds to result in 96.48\% of absolute errors to be less than $\sigma_y/2$ and that is 3.75 times slower than LSMCMC. We also tried setting $\Gamma=1156$ in LEnKF similar to that in LSMCMC algorithm. However, this did not help improving the accuracy of LEnKF as shown in the last row of \autoref{tab:table2}. 

\begin{table}[h!]
\begin{tabular}{|c|c|c|c|c|c|c|}
\hline
\backslashbox{Schemes}{Parameters} &
  $\Gamma$ &
  $r$ &
  $M$ &
  \begin{tabular}[c]{@{}c@{}}$N$ \\ (or $N+N_{\text{burn}}$)\end{tabular} &
  \begin{tabular}[c]{@{}c@{}}Computational\\ time (seconds)\end{tabular} &
  \begin{tabular}[c]{@{}c@{}}\% of absolute\\ errors $<\sigma_y/2$\end{tabular} \\ \hline
LSMCMC & 1156 & -- & 52  & $5000+3000$ & 145.51 & 99.79\% \\ \hline
SMCMC  & --   & -- & 52  & $5000+3000$  & 766.47 & 98.48\% \\ \hline
EnKF   & --   & -- & 1  & 5000        & 485.92 & 93.90\% \\ \hline
LEnKF  & 36   & 130 & 1 & 5000         & 545.13 & 96.48\% \\ \hline
LEnKF  & 1156   & 130 & 1 & 5000         & 1314.89 & 96.47\% \\ \hline
\end{tabular}
\caption{The results of the second test on the linear Gaussian model, where we set $N$ to be the same for all methods.}
\label{tab:table2}
\end{table}

\textbf{Third test:} We choose $N$ in SMCMC, EnKF and LEnKF such that the percentage of absolute errors which are less than $\sigma_y/2$ is equal to that of LSMCMC in \autoref{tab:table1} and we record the computational time in \autoref{tab:table3}. For example, SMCMC took 1934.1 seconds which is 13.3 times slower than LSMCMC. The EnKF and LEnKF are 9.3 and 11.25 times slower than LSMCMC, respectively. It is worth mentioning, as shown in \cite{smcmc}, that when $d$ gets higher, the  cost of SMCMC will become lower than that of EnKF and LEnKF.

\begin{table}[h!]
\begin{tabular}{|c|c|c|c|c|c|c|}
\hline
\backslashbox{Schemes}{Parameters} &
  $\Gamma$ &
  $r$ &
  $M$ &
  \begin{tabular}[c]{@{}c@{}}$N$ \\ (or $N+N_{\text{burn}}$)\end{tabular} &
  \begin{tabular}[c]{@{}c@{}}Computational\\ time (seconds)\end{tabular} &
  \begin{tabular}[c]{@{}c@{}}\% of absolute\\ errors $<\sigma_y/2$\end{tabular} \\ \hline
LSMCMC & 1156 & -- & 52  & $5000+3000$ & 145.51 & 99.79\% \\ \hline
SMCMC  & --   & -- & 52  & $13000+6000$  & 1934.10 & 99.79\% \\ \hline
EnKF   & --  & -- & 1 & 15000        & 1352.89 & 99.79\% \\ \hline
LEnKF  & 36  & 130 & 1 & 15000         & 1636.25 & 99.79\% \\ \hline
\end{tabular}
\caption{The results of the third test on the linear Gaussian model. We keep $\Gamma$ and $r$ as in the first test, but we change $N$ (and $N_{\text{burn}}$) for the other three methods to get the same accuracy as of LSMCMC in the first test.}
\label{tab:table3}
\end{table}

\subsection{Rotating Shallow Water Model with SWOT-Like Observations}
\label{subsec:RSW_SWOT}

We consider the RSWEs given in \eqref{eq:SWE_zeta}-\eqref{eq:SWE_v} below.
\begin{itemize}
\item
The equation for the evolution of the free surface elevation, $\zeta$:
\begin{equation}
\label{eq:SWE_zeta}
\frac{\partial \zeta}{\partial t} + \frac{\partial (\eta u)}{\partial x} + \frac{\partial (\eta v)}{\partial y} = 0
\end{equation}
\item
The equation for the evolution of the horizontal velocity in the $x$-direction, $u$:
\begin{equation}
\label{eq:SWE_u}
\frac{\partial (\eta u)}{\partial t} + \frac{\partial}{\partial x} (\eta u^2 +\tfrac{1}{2}g \eta^2) + \frac{\partial (\eta uv)}{\partial y} = g \eta\frac{\partial H}{\partial x} + f_1 \eta v
\end{equation}
\item
The equation for the evolution of the horizontal velocity in the $y$-direction, $v$:
\begin{equation}
\label{eq:SWE_v}
\frac{\partial (\eta v)}{\partial t} + \frac{\partial (\eta uv)}{\partial x} + \frac{\partial}{\partial y} (\eta v^2 +\tfrac{1}{2}g \eta^2) = g\eta\frac{\partial H}{\partial y} - f_1\eta u
\end{equation}
\end{itemize}
Here, $(x,y) \in  [a_1,b_1]\times [a_2,b_2]$ denotes the spatial domain, $g$ is the gravitational acceleration, $f_1$ is the Coriolis parameter that varies linearly with $y$, $\eta=\zeta+H$ represents the depth of the water, $H$ is the bathymetry, and $\zeta$, $u$, and $v$ are the free surface elevation, and the horizontal $x$- and $y$-velocities, respectively.

For $k\in \{1,\cdots,T\}$, the signal in the SSM is given as follows:
$$
{\mathbf{Z}_t} = \Phi(\mathbf{Z}_{t_{k-1}},t_{k-1};t), \qquad \text{for} ~~ t\in (t_{k-1}, t_k),
$$ 
and at $t_k$
\begin{equation*}
\mathbf{Z}_{t_k} = \Phi(\mathbf{Z}_{t_{k-1}},t_{k-1}; t_k)+ \mathbf{W}_{t_k},
\end{equation*}
where $\mathbf{W}_{t_k}\stackrel{\textrm{\emph{i.i.d.}}}{\sim}\mathcal{N}_d(0,Q_k)$ is an i.i.d.~sequence of $d-$dimensional Gaussian random variables of zero mean and covariance matrix $Q_k$. The observations on the system are for the free surface elevation, and they are arriving as a swath that moves from right to left in a cyclic manner similar to the linear Gaussian example. The map $\Phi(\mathbf{Z}_{t_{k-1}},t_{k-1}; t)$ is the finite-volume (FV) solution of the PDEs \eqref{eq:SWE_zeta}-\eqref{eq:SWE_v} at time $t$ given $\mathbf{Z}_{t_{k-1}}$ as an initial value.

The initial and boundary conditions for this system are realistic data provided by \cite{coper}. We use a FV numerical scheme that comprises a 2-stage Runge-Kutta method combined with a local Lax-Friedrichs scheme (see \cite[Appendix]{smcmc} for more details). The hidden signal is the output
\begin{equation*}
\mathbf{Z}_t = [(\eta^{i}_t)_{1\leq i \leq N_g^xN_g^y }, (u^{i}_t)_{1\leq i \leq N_g^xN_g^y },(v^{i}_t)_{1\leq i \leq N_g^xN_g^y } ]^\top \in \mathbb{R}^{d},
\end{equation*}
where $d=3N_g^xN_g^y$, $\mathbf{Z}_0=\mathbf{z}_0$ is given and $(\eta^i_t)_{1\leq i \leq N_g^x N_g^y}, (u^{i}_t)_{1\leq i \leq N_g^x N_g^y },(v^{i}_t)_{1\leq i \leq N_g^x N_g^y }$ are row vectors obtained from the FV solver. The additive Gaussian noise used here is the same as in \cite{smcmc}. It is designed in a specific way such that it is zero at the boundary of the domain to preserve  boundary conditions. More specifically, for $\eta_{t_k}$, $k\in \mathbb{N}$, we use 
\begin{equation*}
    [\Xi^{\eta}_{t_k}]_{l=1,s=1}^{N_g^y,N_g^x}= \sum_{i=0}^{J-1} \sum_{j=0}^{J-1}\epsilon_{t_k}^{\eta, (i,j)} \sin\left(\frac{\pi j y_l}{b_2-a_2}\right) \sin\left(\frac{\pi i x_s}{b_1-a_1}\right),
\end{equation*}
where $J\in\mathbb{N}$ is a user chosen number of Fourier modes, $\epsilon_{t_k}^{\cdot,(i,j)} \stackrel{\textrm{\emph{i.i.d.}}}{\sim} \mathcal{N}(0,\sigma^2/(i\vee j+1))$, for $i,j \in \{0,\cdots,J-1\}$, where $i\vee j$ here means the maximum of $\{i,j\}$, and $\sigma > 0$ is given. Similarly for $u^{t_k}$ \& $v^{t_k}$ we design $\Xi_{^{t_k}}^u$ \& $\Xi_{^{t_k}}^v$ then vectorize the matrices to get
\begin{equation}
\label{eq:state_noise}
\mathbf{W}_{t_k}=[\text{Vec}(\Xi_{^{t_k}}^\eta)^T, \text{Vec}(\Xi_{^{t_k}}^u)^T,\text{Vec}(\Xi_{^{t_k}}^v)^T]^T \in \mathbb{R}^{d}.
\end{equation}
The covariance matrix $Q_{t_k}\equiv Q\in \mathbb{R}^{d\times d}$ associated with this noise has the form
\begin{align*}
Q = \begin{bmatrix}
Q^\eta & \mathbf{0} & \mathbf{0}\\
\mathbf{0} & Q^u & \mathbf{0}\\
\mathbf{0} & \mathbf{0} & Q^v
\end{bmatrix}
\end{align*}
where $\mathbf{0} \in \mathbb{R}^{N_g^xN_g^y\times N_g^xN_g^y}$ is a zero matrix and $\overline{Q}:= Q^\eta=Q^u=Q^v \in \mathbb{R}^{N_g^xN_g^y\times N_g^xN_g^y}$ has the form
\begin{align*}
\overline{Q} = \begin{bmatrix}
Q_{1,1} & \cdots & Q_{1,N_g^x}\\
\vdots & \vdots & \vdots\\
Q_{N_g^x,1} & \cdots & Q_{N_g^x, N_g^x}
\end{bmatrix}
\end{align*}
where for $1\leq i,j \leq N_g^x$, each block $Q_{i,j}\in \mathbb{R}^{N_g^y\times N_g^y}$, with elements 
$$
Q_{i,j}^{m,n} = \sigma^2\sum_{l=0}^{J-1}\sum_{s=0}^{J-1} S_1^{m,l}S_1^{n,l}S_2^{i,s}S_2^{j,s} \frac{1}{l\vee s + 1}, \qquad m,n = 1, \cdots,N_g^y, 
$$
where
\begin{equation*}
S_1 = \left[\sin\left(\tfrac{\pi j y_l}{b_2-a_2}\right)\right]_{l=1, j = 0}^{N_g^y, J-1} \in \mathbb{R}^{N_g^y \times J}, \qquad S_2 = \left[\sin\left(\tfrac{\pi i x_s}{b_1-a_1}\right)\right]_{s=1, i = 0}^{N_g^x, J-1} \in \mathbb{R}^{N_g^x \times J}.
\end{equation*}
Sampling $\mathbf{W}_{t_k}$ from $\mathcal{N}(\mathbf{0}, Q)$ is simple. We generate three random matrices ${\epsilon}_{t_k}^\eta$, ${\epsilon}_{t_k}^u$ \& ${\epsilon}_{t_k}^v$ of size $J\times J$ each with independent entries, $\epsilon_{t_k}^{.,(i,j)} \stackrel{\textrm{\emph{i.i.d.}}}{\sim} \mathcal{N}(0,\sigma^2/(i\vee j+1))$, for $i,j \in \{0,\cdots,J-1\}$. Then, $\Xi_{t_k}^\eta = S_1 {\epsilon}_{t_k}^\eta S_2^T$, $\Xi_{t_k}^u = S_1 {\epsilon}_{t_k}^u S_2^T$, $\Xi_{t_k}^v = S_1 {\epsilon}_{t_k}^v S_2^T$ and $\mathbf{W}_{t_k}$ is then as in \eqref{eq:state_noise}.
 
\subsubsection*{Simulation Settings and Results}
The simulation domain covers a region of the Atlantic Ocean within the longitude range $[-50^\circ, -30^\circ]$ and latitude range $[17^\circ, 37^\circ]$. Bathymetry data ($H$), sea surface height above the geoid ($\overline{\eta}$), and horizontal velocities ($\overline{u}$ and $\overline{v}$) are sourced from the Copernicus Marine Services \cite{coper}. These variables are provided at a spatial resolution of $1/12$ degree and at an hourly interval, spanning from March 1, 2020, to March 5, 2020. The values of $\overline{\eta}$, $\overline{u}$, and $\overline{v}$ at 00:00 on March 1, 2020, are used as the initial state $\mathbf{z}_0$. For boundary conditions, the values of $\overline{\eta}$, $\overline{u}$, and $\overline{v}$ along the boundaries are extrapolated in time. We set $N_g^x = 121$, $N_g^y=121$, $d = 3N_g^xN_g^y = 4.3923 \times 10^4$, the spatial descritizations $\Delta_x =16.292$ km \& $\Delta_y=18.471$ km, the time-step $\tau_k = 120$ seconds for all $k\in \mathbb{N}$, the number of time steps $T= 1000$ (equivalent to 33.3 hours), the observational-time lag $L=10$, the number of Fourier modes for the noise is $J=8$, the MCMC samples $N=1200$ \& $N_{burn}=500$, the number of independent simulations $M=26$, the state noise hyperparameter $\sigma = 1\times 10^{-3}$. The observations are a noisy set of the water's height in a swath of width $\sim$ 114 km that moves from right to left as in the first example. The observations follow the same linear Gaussian model as in \eqref{eq:numer_obs_model} with $R_k = \sigma_y^2 I_{d_y^k}$, with $\sigma_y=1.45 \times 10^{-2}$. In the LSMCMC scheme we set $\Gamma$ to 1800; this will result in subdomains of size $2\times 4$.

\begin{figure}[h!]
\centering
\includegraphics[scale=0.64]{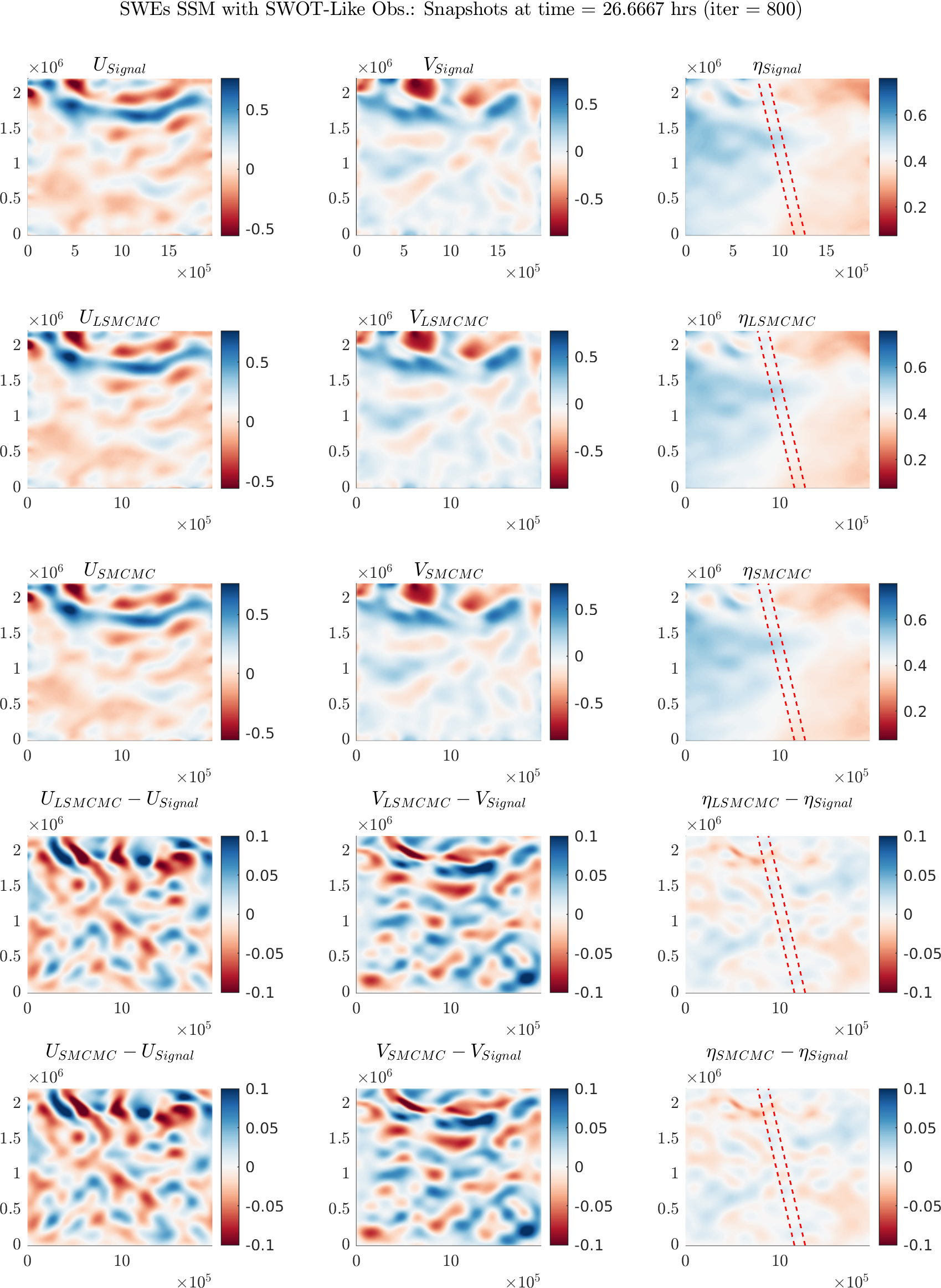}
\caption{SWEs SSM with SWOT-like observations: Snapshot of the LSMCMC and SMCMC filters means at observational time 26.67hrs (that is $k=800$). The first row shows the signal parts $(u,v,\eta)$. The second row shows the mean of the LSMCMC filter for these three parts. The third row shows the mean of the SMCMC filter. Finally, fourth and fifth rows show the difference between the reference and the LSMCMC and SMCMC filters means, respectively. The red dashed lines show the boundary of the observed region of the water's height at $k=800$.}
\label{fig:swe_swath_lsmcmc_smcmc}
\end{figure}

\begin{figure}[h!]
\centering
\includegraphics[scale=0.18]{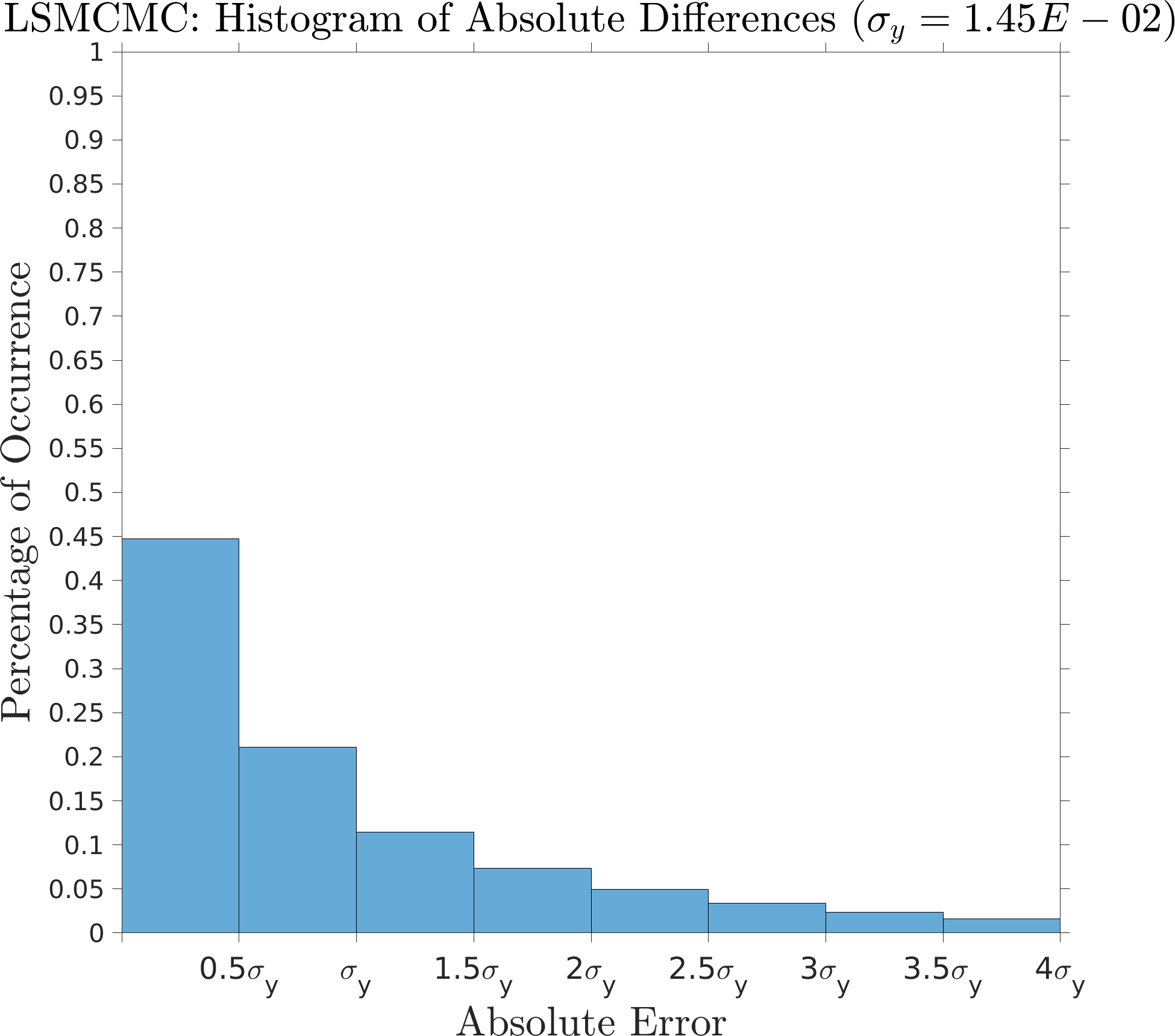} \quad \includegraphics[scale=0.18]{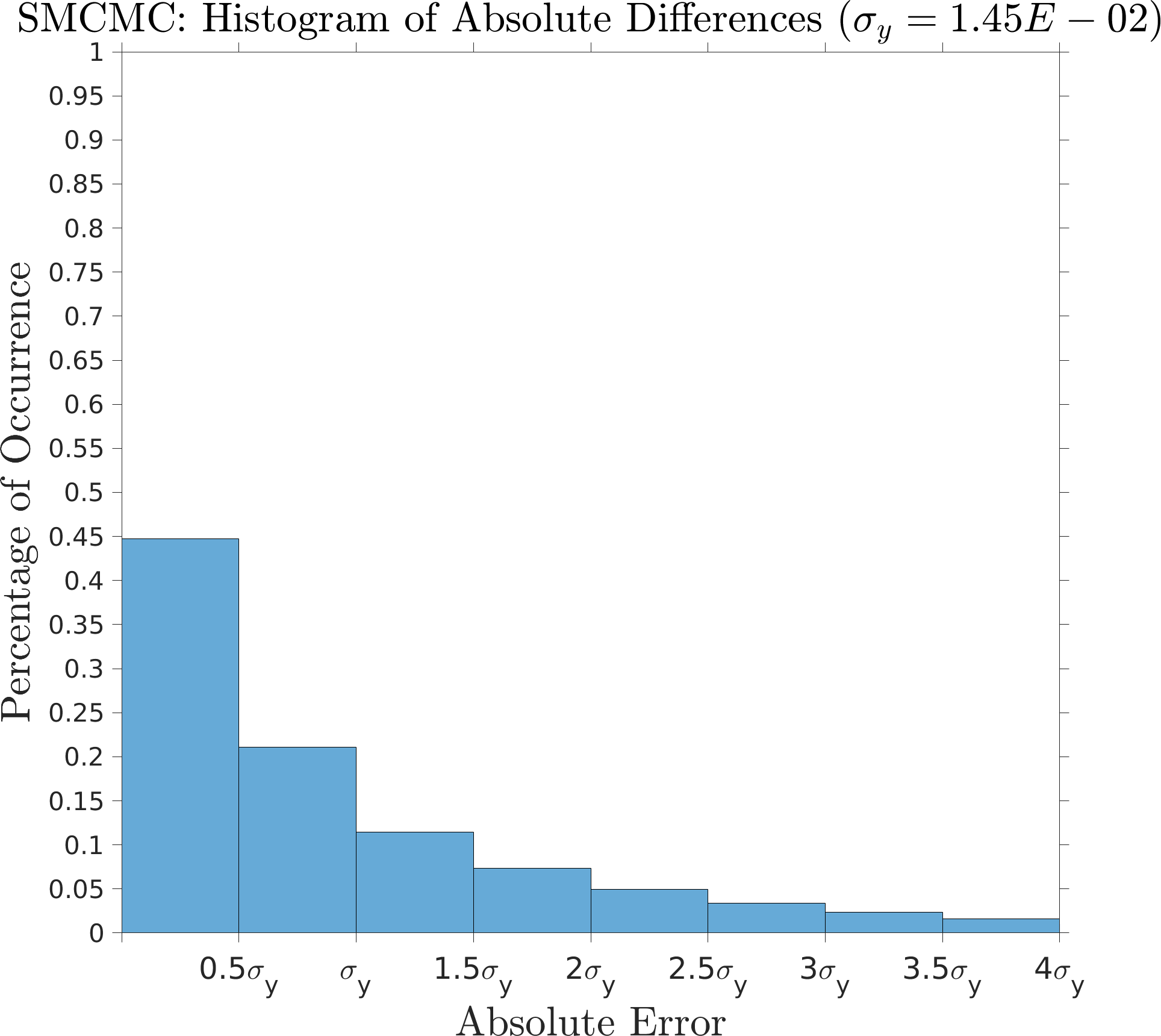}
\caption{SWEs SSM with SWOT-like observations: Histograms of the absolute differences of LSMCMC and SMCMC filters means and the reference signal at every state variable $\mathbf{Z}_{k}^{(i)}$ and time iteration $k$.}
\label{fig:swe_swath_lsmcmc_smcmc_hist}
\end{figure}

We compare LSMCMC versus SMCMC for the same number of samples. We take the signal which the data was generated from to be the reference. In \autoref{fig:swe_swath_lsmcmc_smcmc}, we show a snapshot of the simulations for the LSMCMC and SMCMC filters at $t_k=80$ ($k=800$). The figure shows that for the same number of MCMC and burned-in samples, both filters give essentially the same results. However, the computational time for LSMCMC is 9.5 hours while it is 53 hours for SMCMC; that is, LSMCMC is about 5.6 times faster than SMCMC for similar accuracy as shown in \autoref{fig:swe_swath_lsmcmc_smcmc_hist}. The percentage of absolute differences that are less than $\sigma_y/2$ is exactly 44.76\% for both filters.

\subsection{Rotating Shallow Water Model with Drifters Observations}
\label{subsec:RSW_drifters}

Here we consider the same dynamics as in the previous example except that the observations are now obtained from a set of $N_d$ ocean drifter in the region of interest observing the water horizontal velocities $u$ \& $v$. The drifters data is obtained from \cite{adam1}. Since the drifters data is associated with an uncertainty, we let the locations of the drifters to depend on the hidden signal. That is, it is assumed that the drifters are moving according to the stochastic velocity components of $\mathbf{Z}_k$ that we are trying to infer during the assimilation process. Therefore, one needs to estimate the observational locations at each time of observation. This is a very challenging SSM because the dependence of the spatial locations of observations on the hidden signal yields a long-range dependence (in time) on the signal itself.  

We follow \cite{smcmc} and assume that the unknown ocean velocities will affect the drifter's position through the following relation: For $j\in\{1,\dots,N_d\}$, the drifter's position at time $t$, $\mathbf{X}_t^j$, is given as
\begin{equation*}
d\mathbf{X}_t^{j}  =  \mathcal{H}(\mathbf{X}_t^{j},\mathbf{Z}_t)dt, 
\end{equation*}
for some function $\mathcal{H}:\mathbb{R}^2\times \mathbb{R}^d\to \mathbb{R}^2$ (in our case $\mathcal{H}$ is just the velocity fields). The above relation can be rewritten as
\begin{equation*}
\mathbf{X}_{t_k}^{j} = \mathbf{X}_0^{j} + \int_{0}^{t_k} \mathcal{H}(\mathbf{X}_s^{j},\mathbf{Z}_s) ds, \quad (j,k)\in \{1,\dots,N_d\} \times\mathbb{N}.
\end{equation*}
One can then use the Euler scheme to obtain the following estimate for the location of drifter $j$:
\begin{equation}
\label{eq:estimate_drift_loc}
\overline{\mathbf{X}}_{t_k}^{j} = \overline{\mathbf{X}}_{t_{k-1}}^{j} +  \mathbb{E}\Big[\sum_{l=0}^{L-1}\mathcal{H}(\tilde{\mathbf{X}}^{j}_{t_{k-1}+l\tau_{k}}, \mathbf{Z}_{t_{k-1}+l\tau_{k}})\tau_{k}\Big|(\mathbf{Y}_{t_p})_{p\leq {k-1}},  \mathbf{Z}_{t_{k-1}}= \mathbf{z}_{t_{k-1}}, \tilde{\mathbf{X}}^{j}_{t_{k-1}}=\overline{\mathbf{x}}_{t_{k-1}}^{j} \Big]
\end{equation}
with the expectation taken w.r.t. the pair $\big(\mathbf{Z}_{t_{k-1}+(l-1)\tau_k}, \tilde{\mathbf{X}}^{j}_{t_{k-1}+(l-1)\tau_k}\big)_{1<l\leq L}$ which are obtained via propagating the dynamics of $\mathbf{Z}_t$ for the $N$ samples jointly with the approximation
\begin{equation*}
\tilde{\mathbf{X}}_{t_k}^{j, i} = \tilde{\mathbf{X}}_{t_{k-1}}^{j, i} +  \sum_{l=0}^{L-1}\mathcal{H}(\tilde{\mathbf{X}}_{t_{k-1}+l\tau_{k}}^{j, i}, \mathbf{Z}_{t_{k-1}+l\tau_{k}}^{i})~\tau_{k}, \qquad i \in \{1,\cdots, N\}.
\end{equation*}
Note that at previous times $\{t_{k-1}+(l-1)\tau_k\}_{1<l\leq L}$, we have $N$ paths of $\mathbf{Z}_{t_{k-1}+(l-1)\tau_k}$, hence the need to take the expectation in \eqref{eq:estimate_drift_loc}. Finally, the approximation in \eqref{eq:estimate_drift_loc} allows the drifter to be positioned anywhere within the physical domain. However, due to the discretization required for solving the RSWEs, it is necessary to map the drifter's location from the continuous physical domain to the discretized computational grid. For instance, if the drifter is located within a specific grid cell, one option is to assign the nearest grid point as the observation location. Alternatively, the drifter's observations could be interpreted as influencing the four surrounding grid points.

\subsubsection*{Simulation Settings and Results}
The region of simulation in this example is still a domain of the Atlantic Ocean: $[-51^\circ, -41^\circ]\times [17^\circ, 27^\circ]$. The initial and boundary conditions are obtained from \cite{coper} and extrapolated in time as in the previous example. We set $N_g^x = 121$, $N_g^y=121$, $d = 3N_g^xN_g^y = 4.3923 \times 10^4$, the spatial descritizations $\Delta_x =8.602$ km \& $\Delta_y=9.258$ km, the time-step $\tau_k = 60$ seconds for all $k\in \mathbb{N}$, the number of time steps $T= 1000$ (equivalent to 16.67 hours), the observational-time lag $L=10$, the number of Fourier modes for the noise is $J=8$, the number of independent simulations $M=26$, the state noise hyperparameter $\sigma = 1\times 10^{-3}$. The observations follow the same linear Gaussian model as in \eqref{eq:numer_obs_model} with $R_k = \sigma_y^2 I_{d_y^k}$, with $\sigma_y=1.45 \times 10^{-2}$. In LSMCMC scheme, we run the algorithm with two different values of $\Gamma$, namely, 1440 and 720; this will result in subdomains of size $2\times 5$ and $4\times 5$, respectively.

\begin{figure}[h!]
\centering
\includegraphics[scale=0.64]{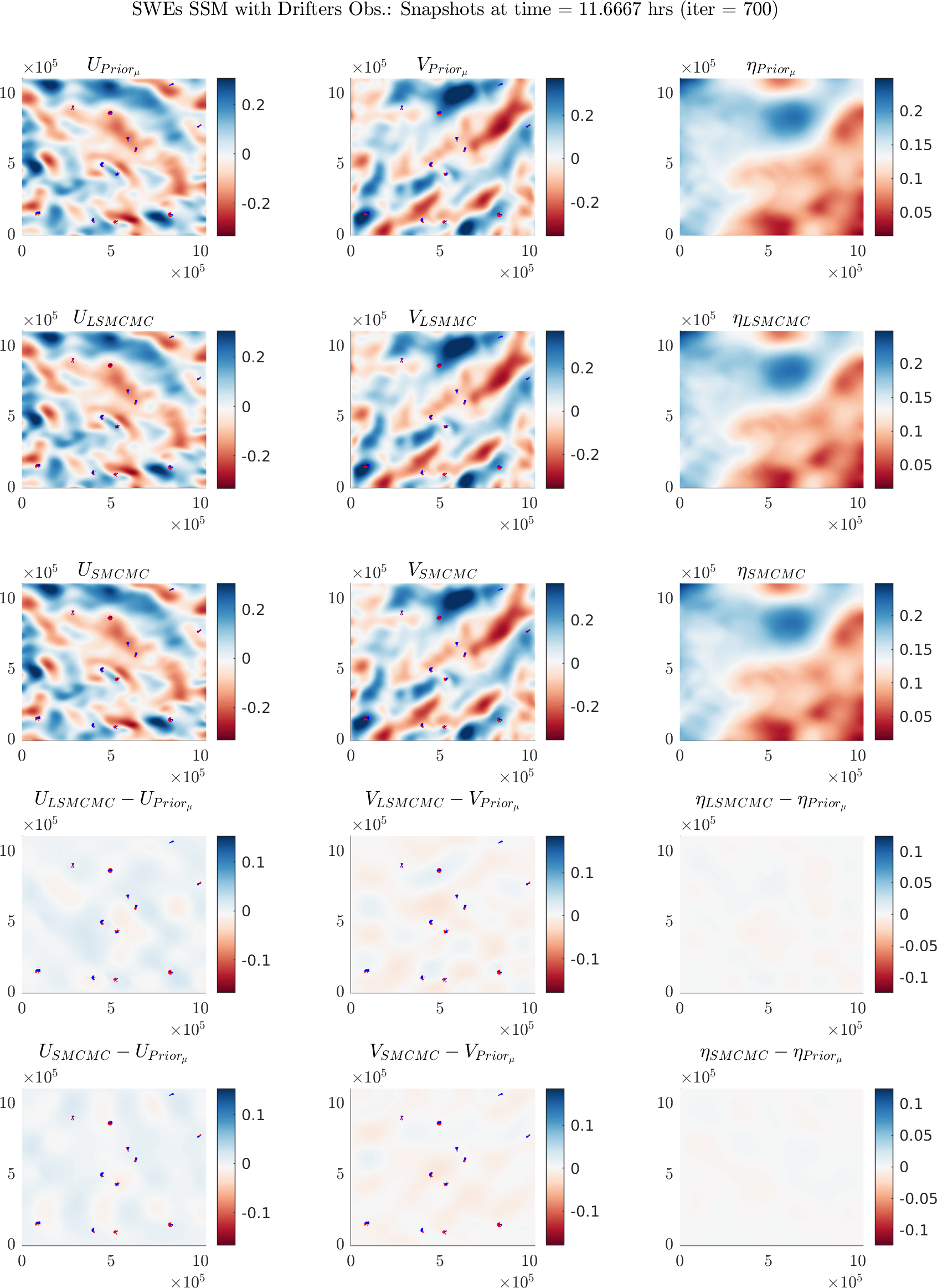}
\caption{SWEs SSM with Drifters Observations: Snapshot of the LSMCMC and SMCMC filters means at observational time 11.67hrs (that is $k=700$). The first row shows the prior parts $(u,v,\eta)$. The second row shows the mean of the LSMCMC filter for these three parts. The third row shows mean of the SMCMC filter. Finally, fourth and fifth rows show the difference between the reference and the LSMCMC and SMCMC filters means, respectively. The blue curves represent the drifter tracks based on the data from \cite{adam1}, while the red curves depict the average of 50 drifter tracks simulated according to the prior. Both the red and blue tracks are shown here solely for illustrative purposes. Refer to \autoref{fig:swe_drift_zoomed} for a closer look.}
\label{fig:swe_drift_lsmcmc_smcmc}
\end{figure}

\begin{figure}[h!]
\centering
\includegraphics[scale=0.23]{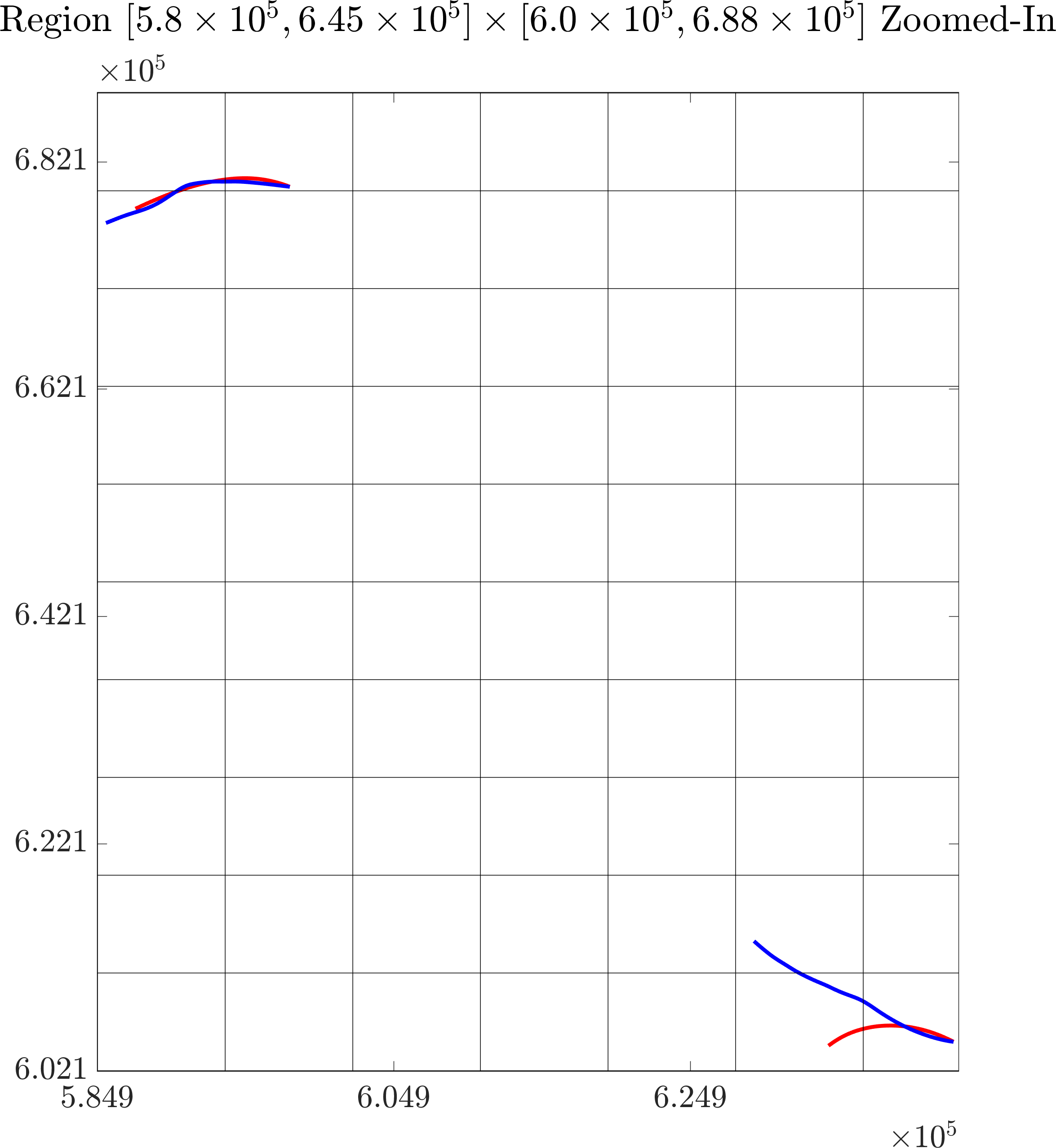} \quad
\includegraphics[scale=0.23]{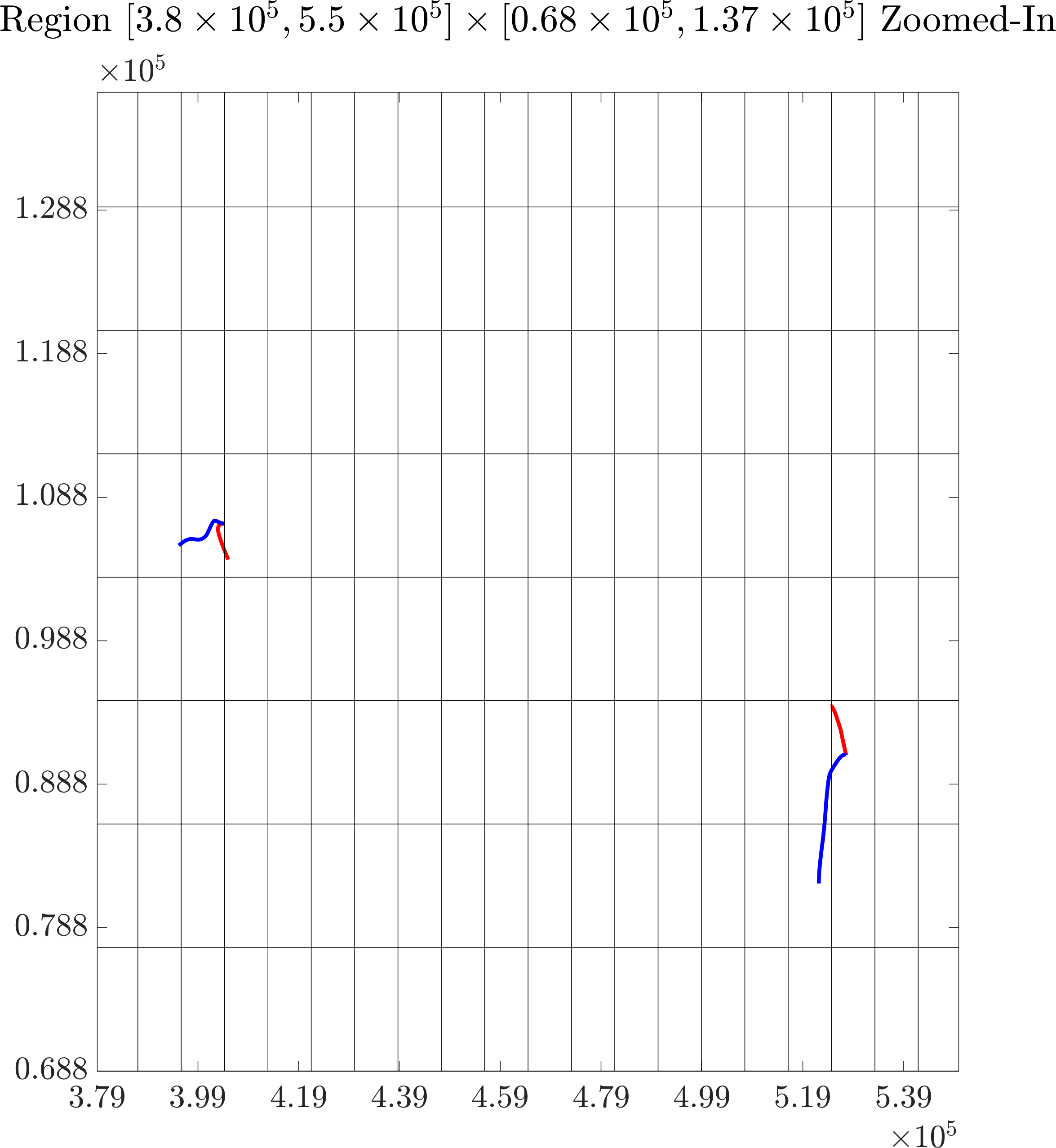}
\caption{SWEs SSM with Drifters Observations: Regions from the previous figure which contain tracks of four different drifters zoomed-in. }
\label{fig:swe_drift_zoomed}
\end{figure}

We compare LSMCMC versus SMCMC for the same number of samples $N$ but different $N_{\text{burn}}$. We set the MCMC samples to be $N=1200$ \& $N_{burn}=200$ in SMCMC (similar to those in \cite{smcmc}) and $N=1200$ \& $N_{burn}=500$ in LSMCMC. Since the observations are realistic, we take the reference to be an average of 50 runs of the forward model which we call the prior. In \autoref{fig:swe_drift_lsmcmc_smcmc}, we show a snapshot of the simulations at $t_k=70$ ($k=700$) when $\Gamma=720$. The figure shows that for the same number of MCMC samples, both filters give comparable results; the percentage of absolute differences between the prior and the LSMCMC filter means that are less than $\sigma_y/2$ is 85.84\% when $\Gamma=1440$ and 88.87\% when $\Gamma=720$, while it is 90.72\% for SMCMC. However, although we use more burn-in samples in LSMCMC, the computational time for LSMCMC when $\Gamma=1440$ and $\Gamma=720$ is 3.3 and 4.67 hours, respectively, while it is 34.3 hours for SMCMC. Thus, with similar accuracy and even larger $N_{\text{burn}}$, LSMCMC in this example is about 10.4 times faster in the first case and 7.35 times faster in the second case.

\section{Conclusion}
\label{sec:conclusion}

In this article, we proposed a novel localization scheme for DA that addresses several challenges in high-dimensional, non-linear, and generally non-Gaussian SSMs. The localization technique focuses on regions with available observations and provides an approximation to the state noise covariance matrix within the framework of SMCMC filtering method. When compared to SMCMC, the localization scheme significantly reduced the computational cost while maintaining high accuracy. Through numerical experiments involving both synthetic data and real oceanographic observations, including swaths of data analogous to those produced by NASA SWOT mission and ocean drifter observations from NOAA, we demonstrated the effectiveness and efficiency of the proposed method. Our approach consistently showed superior performance in terms of computational cost and accuracy. The results also indicate that LSMCMC filter can handle high-dimensional systems with complex dynamics like RSWEs while remaining computationally feasible, marking a step forward in the application of data assimilation techniques to challenging real-world problems. Future work may explore further refinements to the localization strategy and extend the methodology to other types of observational data and dynamical systems. Additional cost reductions may be also achieved when the covariance matrices for state noises are spatial kernels. In such cases, low-rank approximations, such as \cite{kriemann, h2opus}, can compress the matrices to a representation that requires $O(kd)$ storage, where $k$ is a representative rank of the low-rank blocks in these matrices, enabling vector-matrix multiplications of order $O(kd)$. This would ultimately result in an LSMCMC filter with a total complexity of order $O(Nkd')$, where $k\ll d'\ll d$. This is the topic of a current project that we are working on.

\subsection*{Acknowledgements}
This work was supported by KAUST baseline fund.

\subsection*{Data Availability Statement}

The data used in this study for the RSWEs initial values and boundary conditions are openly available from Copernicus Marine Service: Global Ocean Physics Analysis and Forecast at \url{https://doi.org/10.48670/moi-00016}. The drifters' data used in the second RSWEs model are openly available from NOAA at \url{https://doi.org/10.25921/x46c-3620}. The code used to generate the numerical results presented in this study is available on Github at \url{https://github.com/ruzayqat/LSMCMC}.
\appendix
\section{Convergence of SMCMC Filter}

For $k\geq 1$, let $\mathbf{z}_{t_{1:k}} \in A_k \subseteq \mathbb{R}^{dk}$, then the collection of samples $\{\mathbf{Z}_{t_1}^{(1)}, \cdots,\mathbf{Z}_{t_1}^{(N)},\cdots, \mathbf{Z}_{t_{k-1}}^{(1)}, \cdots \mathbf{Z}_{t_{k-1}}^{(N)}\}\in A_{k-1}^N:=A_{k-1}\times  \cdots \times A_{k-1}$. We denote by
\begin{align*}
S_{\mathbf{z}_{t_{1:k}},k-1}^N(d\mathbf{z}_{t_{1:k-1}}) := \frac{1}{N} \sum_{i=1}^N \delta_{\left\{\mathbf{Z}_{t_{k-1}}^{(i)}\right\}}(d\mathbf{z}_{t_{1:k-1}})
\end{align*}
the $N-$empirical measure of the collection of samples $\{\mathbf{Z}_{t_{k-1}}^{(1)},\cdots,\mathbf{Z}_{t_{k-1}}^{(N)}\}$ on the space $A_{k-1}^N$. For a bounded and Borel function $\varphi_k$ on $A_k$, we denote by $\Pi_k(\varphi_k):= \int_{A_k} \varphi(\mathbf{z}_{t_{1:k}}) \Pi_k(\mathbf{z}_{t_{1:k}})d\mathbf{z}_{t_{1:k}}$ the expectation of $\varphi_k$ w.r.t. the smoothing distribution $\Pi_k$. We denote by $K_1$ the MCMC transition kernel of invariant measure 
$$
\Pi_1 := \frac{f_1(\mathbf{z}_0, \mathbf{z}_{t_1})~g_1(\mathbf{z}_{t_1}, \mathbf{y}_{t_1})}{\int_{A_1} f_1(\mathbf{z}_0, \mathbf{z}_{t_1})~g_1(\mathbf{z}_{t_1}, \mathbf{y}_{t_1})~d\mathbf{z}_{t_1}}.
$$
For $k\geq 2$ and $N-$empirical density $S_{\mathbf{z}_{t_{1:k}},k-1}^N$, we denote by $K_{S_{k-1}^N, k}$ the transition kernel with invariant measure 
\begin{align*}
\Pi_k^N(\mathbf{z}_{t_{1:k}}) 
:= \frac{f_k(\mathbf{z}_{t_{k-1}},\mathbf{z}_{t_k}) g_k(\mathbf{z}_{t_k},\mathbf{y}_{t_k}) S_{\mathbf{z}_{t_{1:k}},k-1}^N(\mathbf{z}_{t_{1:k-1}})}{p(\mathbf{y}_{t_k}|\mathbf{y}_{t_{1:k-1}})} = S_{\mathbf{z}_{t_{1:k}},k-1}^N(\mathbf{z}_{t_{1:k-1}}) H_k(\mathbf{z}_{t_{k-1}}, \mathbf{z}_{t_{k}}),
\end{align*}
where 
$$
p(\mathbf{y}_{t_k}|\mathbf{y}_{t_{1:k-1}}) = \int_{A_k} f_k(\mathbf{z}_{t_{k-1}},\mathbf{z}_{t_k}) g_k(\mathbf{z}_{t_k},\mathbf{y}_{t_k}) \Pi_{k-1}(\mathbf{z}_{t_{1:k-1}}) d\mathbf{z}_{t_{1:k}},
$$
and
$$
H_k(\mathbf{z}_{t_{k-1}}, \mathbf{z}_{t_{k}}):= \frac{f_k(\mathbf{z}_{t_{k-1}},\mathbf{z}_{t_k}) g_k(\mathbf{z}_{t_k},\mathbf{y}_{t_k})}{p(\mathbf{y}_{t_k}|\mathbf{y}_{t_{1:k-1}})}.
$$
Then we have the following assumption.
\begin{ass}
\label{ass:A}
\begin{enumerate}
\item $[$Uniform ergodicity of $K_1$.$]$ There exist $\beta_1 \in (0,1)$ such that for any $\mathbf{z}_{t_1} \in A_1$, 
$$
K_1(\mathbf{z}_{t_1}, \cdot) \geq \beta_1 \Pi_1(\cdot).
$$
\item $[$Uniform ergodicity of $K_{S_{k-1}^N, k}$.$]$ There exist $\beta_k \in (0,1)$ such that for any $\mathbf{z}_{t_{1:k}} \in A_k$ and any collection of points $\{\mathbf{Z}_{t_{k-1}}^{(1)}, \cdots, \mathbf{Z}_{t_{k-1}}^{(N)}\} \in A_{k-1}^N\setminus A_{k-2}^N$,
$$
K_{S_{k-1}^N, k}(\mathbf{z}_{t_{1:k}}, \cdot) \geq \beta_k \Pi_k^N(\cdot).
$$
\item $[H_k(\mathbf{z}_{t_{k-1}}, \cdot) \in L_1(A_k\setminus A_{k-1})$$]$. That is, for any $k\geq 2$,
$$
\sup_{\mathbf{z}_{t_{1:k-1}} \in A_{k-1}}\int_{A_k\setminus A_{k-1}} |H_k(\mathbf{z}_{t_{k-1}}, \mathbf{z}_{t_k})| d\mathbf{z}_{t_k} < \infty.
$$
\end{enumerate}
\end{ass}

The following result assures the almost sure convergence of the SMCMC filter.
\begin{prop}
\label{prop:1}
Under \autoref{ass:A}, for any $k\geq 1$, data $\mathbf{Y}_{t_{1:k}}$, and $p\geq 1$, there exists a constant $C_{p,k}(\beta_k,\mathbf{Y}_{t_{1:k}}) <\infty$ such that for any $\varphi_k$ bounded and Borel function on $A_k$:
\begin{equation*}
\mathbb{E}_{\mathbf{z}_{t_1}^{(1)}} \left[\left|\frac{1}{N} \sum_{i=1}^N \varphi_k(\mathbf{Z}_{t_{1:k}}^{(i)}) - \Pi_k(\varphi_k)\right|^p \Bigg|\mathbf{Y}_{t_{1:k}} \right]^{1/p} \leq \frac{C_{p,k}(\beta_k,\mathbf{Y}_{t_{1:k}}) \|\varphi_k\|_\infty}{\sqrt{N}}.
\end{equation*}
\end{prop}
The proof can be found in \cite{martin} with some minor changes. The proof uses Poisson equation and induction on $k$ and basically  relies on results of control of adaptive MCMC chains from \cite{andrieu}.

\section{The Pseudocodes for SMCMC and LEnKF}
The original algorithm of SMCMC filter is presented below.

\setcounter{algorithm}{0}
\renewcommand{\thealgorithm}{B.\arabic{algorithm}}
\begin{flushleft}
\captionsetup[algorithm]{style=algori}
\captionof{algorithm}{Pseudocode for SMCMC filtering method for $T$ observational time steps.}
\label{alg:smcmc}

 \textbf{Input:} The initial state $\breve{\mathbf{Z}}_0=\mathbf{z}_0$, the observations $\{\mathbf{Y}_{t_k} = \mathbf{y}_{t_k}\}_{k\geq 1}$, and the number of time discretization steps (observational time-lag) $L$. Set $\tau_k=(t_k-t_{k-1})/L$. 

\begin{enumerate}
\item Initialize: For $l=0,\cdots,L-1$ compute $\breve{\mathbf{Z}}_{(l+1)\tau_1} = \Phi(\breve{\mathbf{Z}}_{l\tau_1},l\tau_1;(l+1)\tau_1)$. Compute $\tilde{\mathbf{Z}}_{t_1}:=\breve{\mathbf{Z}}_{t_1}+ \mathbf{W}_{t_1}$, where $\mathbf{W}_{t_1}\sim \mathcal{N}_d(0,Q_1)$. \textcolor{black}{Run a standard RWM initialised at $\tilde{\mathbf{Z}}_{t_1}$} to generate $N$ samples $\{\mathbf{Z}_{t_1}^{(i)}\}_{i=1}^N$ from $\pi_1$ in \eqref{eq:pi_1}. Set $\widehat{\pi}_{1}^N(\varphi) \leftarrow \frac{1}{N} \sum_{i=1}^N \varphi(\mathbf{Z}_{t_1}^{(i)})$.

\item For $k=2,\ldots,T$: 
\begin{enumerate}
\item For $i=1,\cdots, N$: compute 
$$\breve{\mathbf{Z}}_{(l+1)\tau_k+t_{k-1}}^{(i)} = \Phi(\breve{\mathbf{Z}}_{l\tau_k+t_{k-1}}^{(i)},l\tau_k+t_{k-1};(l+1)\tau_k+t_{k-1}),$$ 
where $0\leq l \leq L-1$ and $\breve{\mathbf{Z}}^{(i)}_{t_{k-1}}= \mathbf{Z}^{(i)}_{t_{k-1}}$.  
 
\item \textcolor{black}{Run \autoref{alg:rwm} to return samples $\{\mathbf{Z}_{t_k}^{(i)}\}_{i=1}^N$}. Set $\widehat{\pi}_{k}^N(\varphi) \leftarrow \frac{1}{N} \sum_{i=1}^N \varphi(\mathbf{Z}_{t_k}^{(i)})$. 
\end{enumerate}
\end{enumerate}
\textbf{Output:} Return $\{\widehat{\pi}_k^N(\varphi)\}_{k\in\{1,\cdots,T\}}$.

\vspace{-0.1cm}
\hrulefill
\vspace{0.2cm}
\end{flushleft}
%
%
\begin{flushleft}
\captionsetup[algorithm]{style=algori}
\captionof{algorithm}{Pseudocode for RWM to sample from $\pi_k^{N}(z_{t_k},j)$}
\label{alg:rwm}

\textbf{Initialization:} Sample the auxiliary variable $j_0 \sim p(j)$ (uniformly). Set $\mathbf{Z}_{t_k}^{(0)} \longleftarrow \breve{\mathbf{Z}}^{(j_0)}_{t_k}+ \mathbf{W}_{t_k}$, where $\mathbf{W}_{t_k}\sim \mathcal{N}_d(0,Q_k)$.  Compute
$
\pi_{\text{old}} \longleftarrow  g_k(\mathbf{Z}_{t_k}^{(0)},\mathbf{y}_{t_k}) ~f_k(\mathbf{Z}_{t_{k-1}}^{(j_0)},\mathbf{Z}_{t_k}^{(0)}).$ 

\bigskip 

For $i ={1,\ldots,N+N_{\text{burn}}}$:
\begin{enumerate}

\item Compute proposal for the:
\begin{itemize}
\item  state: $\mathbf{Z}'_{t_k} \longleftarrow \mathbf{Z}_{t_k}^{(i-1)} + \mathbf{W}'$, where $\mathbf{W}' \sim \mathcal{N}_d(0,{Q}')$; 
\item auxiliary variable: $j' = \left\{\begin{array}{ll}
		j_{i-1}-1 & \textrm{if}~j_{i-1}\notin\{1,N\}~\textrm{{w.p.}}~q\\
		j_{i-1} &  \textrm{if}~j_{i-1}\notin\{1,N\}~\textrm{{w.p.}}~1-2q\\
		j_{i-1}+1 & \textrm{if}~j_{i-1}\notin\{1,N\}~ \textrm{{w.p.}}~q\\
j_{i-1}+1 & \textrm{if}~j_{i-1}=1\\
j_{i-1}-1 & \textrm{if}~j_{i-1}=N
		\end{array}\right.$, where $q\in(0,\frac{1}{2}]$.
\end{itemize}
\item Compute 
$
\pi_{\text{new}} \longleftarrow  \left\{\begin{array}{ll}
		g_k(\mathbf{Z}_{t_k}',\mathbf{y}_{t_k}) ~f_k(\mathbf{Z}_{t_{k-1}}^{(j')},\mathbf{Z}_{t_k}') & \textrm{if}~j_{i-1}\notin\{1,N\}\\
		g_k(\mathbf{Z}_{t_k}',\mathbf{y}_{t_k}) ~f_k(\mathbf{Z}_{t_{k-1}}^{(j')},\mathbf{Z}_{t_k}') ~q &  \textrm{if}~j_{i-1}\in\{1,N\}
		\end{array}\right.  .$ 
\item  Compute $\alpha = \min\{1, \pi_{\text{new}}/\pi_{\text{old}}\}$. Sample $u \sim \mathcal{U}[0,1]$. \begin{itemize}
\item  If $ u<\alpha$ set $\mathbf{Z}_{t_k}^{(i)}\longleftarrow \mathbf{Z}_{t_k}'$, $j_i\longleftarrow j'$ and $\pi_{\text{old}}\longleftarrow\pi_{\text{new}}$.
\item Else set $\mathbf{Z}_{t_k}^{(i)}\longleftarrow \mathbf{Z}_{t_k}^{(i-1)}$ and $j_{i} \longleftarrow j_{i-1}.$
\end{itemize}
\end{enumerate}
\textbf{Output:} Return the sequence of $N$ final samples $\{\mathbf{Z}_{t_k}^{(i)}\}_{i=N_{\text{burn}}+1}^N$.

\vspace{-0.1cm}
\hrulefill
\vspace{0.2cm}
\end{flushleft}

We also provide the LEnKF scheme that is tested in \autoref{subsec:linear_model}. We note that we use the Gaspari--Cohn function defined on $\mathbb{R}_+$ for the weight computations as follows:

\begin{align*}
\renewcommand*{\arraystretch}{1.3}
S(x) = \left\{\begin{array}{cl}
-\frac{1}{4}x^5 + \frac{1}{2}x^4 + \frac{5}{8} x^3 -\frac{5}{3}x^2 + 1 & \text{if } x<1\\
\frac{1}{12}x^5 -\frac{1}{2}x^4 + \frac{5}{8} x^3 + \frac{5}{3} x^2 - 5x + 4 - \frac{2}{3x} & \text{if } 1 \leq x \leq 2\\
0 & \text{if } x>2
\end{array} \right. .
\end{align*}

\begin{flushleft}
\captionsetup[algorithm]{style=algori}
\captionof{algorithm}{Localization Scheme of EnKF}
\label{alg:LEnKF}

\textbf{Input:} The domain partition $\mathsf{G}=\bigcup_{i=1}^{\Gamma} G_i$, the grid points $(x,y)$-coordinates $\mathbf{x}_g$ \& $\mathbf{y}_g$, the localization length scale factor $r$, a minimum weight $w_0$ (we use $w_0=1.0E-10$), the ensemble size $N$, and the data $\{\mathbf{Y}_{t_k}=\mathbf{y}_{t_k}\}_{k=1}^{T}$.

\begin{enumerate}

\item Initialize: $\mathbf{Z}_a$ and $\mathbf{Z}_f$ are matrices of size $d\times N$. For $i=1,\cdots,N$ set $\mathbf{Z}_a[:,i] \longleftarrow \mathbf{z}_0$.

\item For $k = 1, \cdots, T$:
\begin{enumerate}
	\item For $i=1, \cdots,N$: $\mathbf{Z}_f[:,i] \longleftarrow A \mathbf{Z}_a[:,i] + \sigma_z \mathbf{W}_i$, \quad $\mathbf{W}_i  \stackrel{\textrm{i.i.d.}}{\sim} \mathcal{N}_d(0,I_d)$. \#\textit{this can be vectorized}
	\item Set $d_y^k \longleftarrow \text{size}[\mathbf{y}_{t_k}]$ and return the $(x,y)$-coordinates of the observations on the grid $\mathbf{x}_o^k$ \& $\mathbf{y}_o^k$. Generate the the matrix $C\in \mathbb{R}^{d_y^k\times d}$ as in \eqref{eq:matrix_C}. Set $\mathbf{v} \longleftarrow \sigma_y^2 \mathbf{1}$ where $\mathbf{1}\in \mathbb{R}^{d_y^k}$ a vector of ones.
	\item Compute the distance between every grid point and every observation location, then return the weights for all grid points as follows. For $i=1,\cdots,d$ and $j=1,\cdots,d_y^k$:
\begin{align*}
\text{Weights}[i,j] = S\left(\frac{\sqrt{(\mathbf{x}_g[i] -\mathbf{x}_o^k[j])^2 + (\mathbf{y}_g[i] - \mathbf{y}_o^k[j])^2}}{r} \right)\\
 \quad \text{\textit{\#can get rid of both for-loops by vectorizing}}
\end{align*}
\item The following for-loop can be done in parallel. Return the grid points in every subdomain. For $i=1,\cdots,\Gamma$:
\begin{itemize}
\item Locate the grid points labels $L_i$ inside subdomain $G_i$; $L_i \subset \{1,\cdots,d\}$. Set $d_i \longleftarrow \text{size}[L_i]$. 
\item Set $W_i \longleftarrow \text{Mean}(\text{Weights}[L_i,:])$, where the mean is taken over the rows of indices in $L_i$ (i.e., $W_i \in \mathbb{R}^{d_y^k}$). Then return the observations labels $O_i$ where $w_i > w_0$; $O_i \subset \{1,\cdots, d_y^k\}$. Set $d_y^{k,i} \longleftarrow \text{size}[O_i]$.
\item Set $C_i \longleftarrow C[O_i,:]$, $\mathbf{y}_{t_k}^i \longleftarrow \mathbf{y}_{t_k}[O_i]$ and $R_k^i \longleftarrow \text{Diag}(\mathbf{v}~/.~ W_i[O_i])$ where $/.$ refers to the element-wise division. Set $\mathbf{Z}_f^i \longleftarrow \mathbf{Z}_f[L_i,:]$.
\item Local update that returns a local updated analysis submatrix $\mathbf{Z}_a^i\in \mathbb{R}^{d_i\times N}$: One performs an analysis update using either EnKF or EnKF with Sherman-Morrison-Woodbury formula using the localized objects $d_i, d_y^{k,i}$, $C_i$, $\mathbf{y}_{t_k}^i$, $\mathbf{Z}_f^i$ and $R_k^i$.  
\item Set $\mathbf{Z}_a[L_i,:] \longleftarrow \mathbf{Z}_a^i$
\end{itemize}
\end{enumerate}
\end{enumerate}

\textbf{Output:} Return $\{\mathbb{E} (\mathbf{Z}_a^k)\}_{k\in\{1,\cdots,T\}}$ where the mean is taken over the ensemble.

\vspace{-0.1cm}
\hrulefill
\vspace{0.2cm}
\end{flushleft}

\end{document}